\documentclass[sigconf]{acmart}

\usepackage{epsfig,endnotes,subcaption}
\usepackage{xspace}
\usepackage{amsmath}
\usepackage{multirow}
\usepackage{xcolor}
\usepackage{pifont}
\usepackage{todonotes}
\usepackage{hyperref}
\usepackage[normalem]{ulem}
\usepackage{balance}
\def\approach{{\sc Tiresias}\xspace}

\hypersetup{%
  pdftitle={Tiresias},
  pdfauthor={Yun Shen},
  pdfkeywords={{Intrusion Prevention}, {Security}, {Malware}},
  pdfsubject={},
  colorlinks=true,
  linkcolor=linkcol,
  citecolor=citecol,
  urlcolor=urlcol,
  pdfstartview=FitB}

\definecolor{linkcol}{rgb}{0,0,0.5}
\definecolor{citecol}{rgb}{0,0.5,0.3}
\definecolor{urlcol}{rgb}{0.3,0,0}

\begin{document}

\date{}

\newcommand{\etal}{\textit{et al}.}
\newcommand{\eg}{e.g., \xspace}
\newcommand{\ie}{i.e., \xspace}

\acmDOI{10.1145/3243734.3243811}
\copyrightyear{2018} 
\acmYear{2018} 
\setcopyright{acmlicensed}
\acmConference[CCS '18]{2018 ACM SIGSAC Conference on Computer and Communications Security}{October 15--19, 2018}{Toronto, ON, Canada}
\acmBooktitle{2018 ACM SIGSAC Conference on Computer and Communications Security (CCS '18), October 15--19, 2018, Toronto, ON, Canada}
\acmPrice{15.00}


\fancyhead{}

\title{\approach: Predicting Security Events Through Deep Learning}
 \author{Yun Shen$^\vardiamondsuit$, Enrico Mariconti$^\clubsuit$, Pierre-Antoine Vervier$^\vardiamondsuit$, and Gianluca Stringhini$^{\spadesuit\clubsuit}$}
 \affiliation{$^\vardiamondsuit$Symantec Research Labs, $^\clubsuit$University College London, $^\spadesuit$Boston University}
 \affiliation{\{yun\_shen,pierre-antoine\_vervier\}@symantec.com, e.mariconti@cs.ucl.ac.uk, gian@bu.edu}

\begin{abstract}
With the increased complexity of modern computer attacks, there is a need for defenders not only to \emph{detect} malicious activity as it happens, but also to \emph{predict} the specific steps that will be taken by an adversary when performing an attack.
However this is still an open research problem, and previous research in predicting malicious events only looked at binary outcomes (\eg whether an attack would happen or not), but not at the specific steps that an attacker would undertake.
To fill this gap we present \approach, a system that leverages Recurrent Neural Networks (RNNs) to predict future events on a machine, based on previous observations.
We test \approach on a dataset of 3.4 billion security events collected from a commercial intrusion prevention system, and show that our approach is effective in predicting the next event that will occur on a machine with a precision of up to 0.93. 
We also show that the models learned by \approach are reasonably stable over time, and provide a mechanism that can identify sudden drops in precision and trigger a retraining of the system. Finally, we show that the long-term memory typical of RNNs is key in performing event prediction, rendering simpler methods not up to the task.
\end{abstract}

\renewcommand\footnotetextcopyrightpermission[1]{}
\maketitle

\thispagestyle{empty}

\section{Introduction} \label{sec:introduction}

The techniques used by adversaries to attack computer systems and networks have reached an unprecedented sophistication.
Attackers use multiple steps to reach their targets~\cite{chen2014study,stringhini2015ain} and these steps are of heterogeneous nature, from sending spearphishing emails containing malicious attachments~\cite{le2014look}, to performing drive-by download attacks that exploit vulnerabilities in Web browsers~\cite{cova2010detection,provos2007ghost}, to privilege escalation exploits~\cite{provos2003preventing}.
After the compromise, miscreants can monetize their malware infections in a number of ways, from remotely controlling the infected computers to stealing sensitive information~\cite{farinholt2017catch,stone2009your} to encrypting the victim's data and holding it hostage~\cite{kharraz2015cutting,kolodenker2017paybreak}. 

Traditionally, the computer security  community has focused on \emph{detecting} attacks by using a number of statistical techniques~\cite{cova2010detection,ho2017detecting,gu2007bothunter,kruegel2003anomaly,stringhini2017marmite,warrender1999detecting}.
While this is inherently an arms race, detection systems provide the foundation for network and system defense, and are therefore very important in the fight against network attacks.
More recently, the attention of the community switched to \emph{predicting} malicious events.
Recent work focused on predicting whether a data breach would happen~\cite{liu2015cloudy}, whether hosts would get infected with malware~\cite{leyla2017riskteller}, whether a vulnerability would start being exploited in the wild~\cite{sabottke2015vulnerability}, and whether a website would be compromised in the future~\cite{soska2014automatically}.
These approaches learn the attack history from previous events (\eg historical compromise data) and use the acquired knowledge to predict future ones.
Being able to predict whether an attack will happen or not can be useful in a number of ways. 
This can for example inform law enforcement on the next target that will be chosen by cybercriminals, enable cyber insurance underwriters to assess a company's future security posture, or assist website administrators to prioritize patching tasks.  

While useful to predictively assess risk in systems and organizations, existing prediction systems have two main limitations.
First, they only focus on predicting events with a binary outcome (\eg whether an attack or a data breach will happen) but do not provide any insights on the techniques and the modus operandi that will be followed by attackers. 
Second, these systems need labeled data to build a model and be able to make a decision, and it is not always easy to obtain such labeled data at a scale that allows to train an accurate model. 
One additional issue is that attackers operate changes to their modus operandi over time, and therefore both feature engineering and binary detection systems themselves need to be updated and retrained to keep performing accurate detections~\cite{mariconti2016mamadroid}.

In this paper, we go beyond predicting a binary outcome on security events, but we rather want to predict the \emph{exact actions} that will be taken by an attacker when performing a computer attack.
To this end, we leverage recent advances in the area of deep learning to develop \approach, a system that learns from past system events and can predict the specific event that will happen next. 
\approach can provide much more precise predictive information compared to previous work, allowing companies to deploy truly proactive countermeasures based on the predicted information. 
For example, this system could predict which particular CVE will be used by an attacker when mounting a multi-step attack against a server, or assess the potential severity of an attack by only looking at its early steps.

\begin{figure*}[t]
	\centering
	\includegraphics[width=\linewidth]{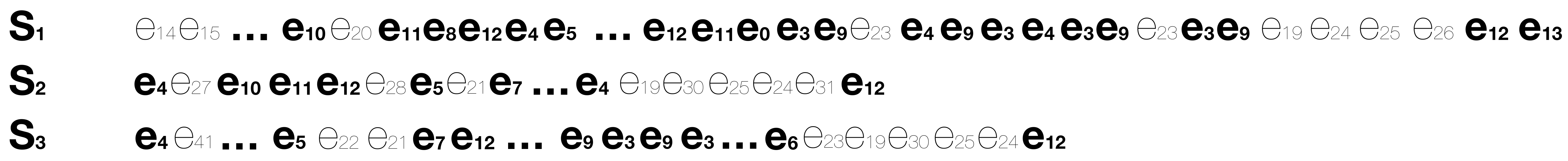}
	\caption{Three endpoints undergoing a coordinated attack. \{$e_0, ..., e_{13}$\} are events involved in the coordinated attack and highlighted in bold.}
	\label{fig:endpoint_sequences}
	\vspace{-0.3cm}
\end{figure*}

We test \approach on a dataset of 3.4 billion security events collected by a commercial intrusion prevention system from a daily population of 740k machines over a period of 27 days.
We show that \approach is effective in predicting the specific malicious event that will occur with a precision of up to 0.93.
We also show that the models trained by \approach are fairly stable, with the detection performance changing slowly. We however identify sudden drops in prediction precision, due to unexpected changes in the observed attacks and systems (e.g., new vulnerabilities being discovered or old ones being patched). In these cases, we show that \approach can automatically identify the issue and trigger a retraining of the system.
We demonstrate that the long-term memory provided by Recurrent Neural Networks is key to performing an accurate prediction when dealing with complex multi-step attacks and the noise generated by legitimate or non-related malicious events in the wild; in fact, we show that \approach clearly outperforms simpler systems that are only based on short-term memory (\eg Markov Chains).

\section{Background and Motivation} \label{sec:motivation}
To illustrate the complexity of keeping track of events across different machines, consider the real-world example in Figure~\ref{fig:endpoint_sequences}. We show three endpoints undergoing a coordinated attack to Apache Web servers (see Section~\ref{sec:case_studies} for the detailed case study), where \{$e_0, ..., e_{13}$\} are events involved in this attack and are highlighted in bold. This coordinated attack consists of three parts: (i) run reconnaissance tasks if port \texttt{80/tcp} (HTTP) is open (\eg $e_4$ is default credential login, $e_5$ is Web server directory traversal), (ii) trigger a list of exploits against the Web application framework Struts (\eg $e_8$ is an exploit relating to Apache Struts CVE-2017-12611, $e_{11}$ is an attempt to use Apache Struts CVE-2017-5638, $e_{13}$ tries to exploit Apache Struts CVE-2017-9805, etc.) and (iii) execute a list of exploits against other possible applications running on the system (\eg $e_2$ exploits Wordpress arbitrary file download, $e_9$ targets Joomla local file inclusion vulnerability, etc). 

The first challenge that we can immediately notice in Figure~\ref{fig:endpoint_sequences} is that even though those three endpoints are going through the same type of attack, there is not an obvious pattern in which a certain event $e_i$ would follow or precede another event $e_j$ given $e_i, e_j \in \{e_0, ..., e_{13}\}$. 
For example, $e_{12}$ (Malicious OGNL Expression Upload) can be followed by $e_4$ (HTTP Apache Tomcat UTF-8 Dir Traversal) and $e_{13}$ (Apache Struts CVE-2017-9805) in $s_1$, yet, it is followed by $e_7$ (Wordpress Arbitrary File Download) and $e_{11}$ (Apache Struts CVE-2017-5638) in $s_2$. 

The second challenge is that the endpoints may observe other security events not relating to the coordinated attack. For example, in $s_3$, we can observe a subsequence \{$e_4, e_{41}, ..., e_{5}, e_{22}, e_{21}, e_7$\} in which $e_5$ is followed by a number of unrelated events including $e_{41}$ (WifiCam Authentication Bypass) before reaching $e_5$. Note that the other noisy events are omitted for the sake of clarity. Between $e_5$ and $e_7$, there were two other noisy event $e_{22}$ (Novell ZENWorks Asset Management) and $e_{21}$ (ColdFusion Remote Code Exec). 

More interestingly, some of these endpoints may potentially observe different attacks from various adversary groups happening at the same time. For example, we observe \{$e_9, e_{19}, e_{24}, e_{25}, e_{26}, e_{12}$\} in $s_1$, \{$e_4, e_{19}, e_{30}, e_{25}, e_{24}, e_{31}, e_{12}$\} in $s_2$, and \{$e_6, e_{23}, e_{19}, e_{30}, e_{25}, e_{24}, e_{12}$\} in $s_3$. It is possible that $e_{19}$ (SMB Validate Provider Callback CVE-2009-3103), $e_{25}$ (SMB Double Pulsar Ping),  and $e_{24}$ (Microsoft SMB MS17-010 Disclosure Attempt) could be part of another coordinated attack. Facing these challenges, it is desirable to have a predictive model that is able to understand noisy events, recognize multiple attacks given different contexts in a given endpoint, and correctly forecast the upcoming security event. This is a more complex and difficult task than detecting each malicious event passively. 

\noindent \textbf{Problem Formulation.} We formalize our security event prediction problem as follows. A security event $e_{j} \in E$ is a timestamped observation recorded at timestamp $j$, where $E$ denotes the set of all unique events and $|E|$ denotes the size of $E$. A security event sequence observed in an endpoint $s_i$ is a sequence of events ordered by their observation time, $s_i=\{e^{(i)}_{1}, e^{(i)}_{2}, ..., e^{(i)}_{n}\}$. We define the to-be-predicted event as \emph{target event}, denoted as $e_{tgt}$. Each target event $e_{tgt}$ is associated with a number of already observed security events, denoted as $l$. The problem is to learn a sequence prediction function $f: \{e_{1}, e_{2}, ..., e_{l}\} \rightarrow e_{tgt}$ that accepts a \emph{variable-length} input sequence $\{e_{1}, e_{2}, ..., e_{l}\}$ and predicts the target event $e_{tgt}$ for a given system. Note that our problem definition is a significant departure from previous approaches that accept only fixed-length input sequences. We believe that a predictive system should be capable of understanding and making predictions given \emph{variable-length} event sequences as the contexts,  hence our problem definition is a better formulation inline with real world scenarios. 

\section{Methodology} \label{sec:workflow}
In this section we describe the system architecture and the technical details behind \approach. 

\subsection{Architecture Overview} \label{sec:architecture}

\begin{figure*}[h]
	\centering
	\includegraphics[width=0.8\linewidth]{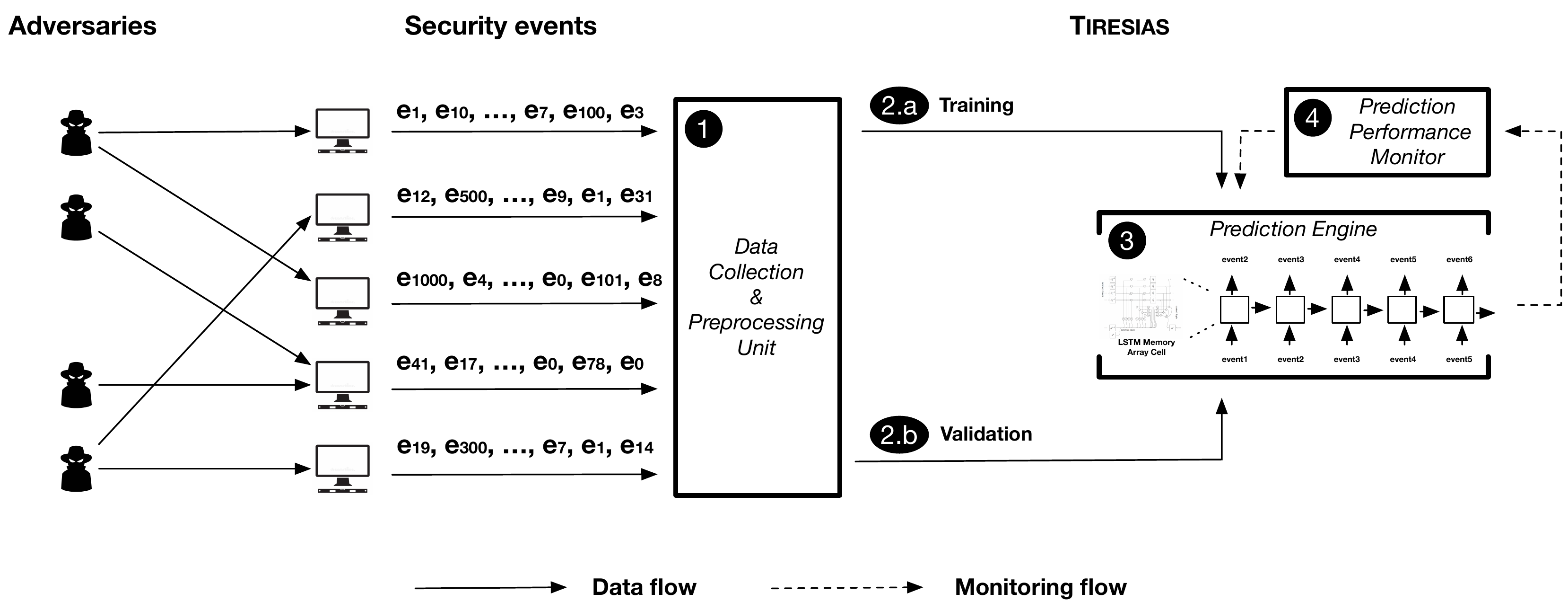}
	\caption{\approach collects security events from machines that have installed an intrusion protection product. The sequential events from these machines are collected, preprocessed and then used to build and validate \approach' predictive model. The optimal model is then used in operations and its performance is monitored to ensure steadily high prediction accuracy.}
	\label{fig:workflow}
	\vspace{-0.3cm}
\end{figure*}

The architecture and workflow of \approach is depicted in Figure~\ref{fig:workflow}. Its operation consists of four phases: \ding{182} data collection and preprocessing, \ding{183} model training \& validation, \ding{184} security event prediction, and \ding{185} prediction performance monitoring.

\noindent \textbf{Data collection and preprocessing (\ding{182}).} \approach takes as input a sequence of security events generated by endpoints (\eg computers that installed a security program). The goal of the data collection and preprocessing module is to prepare both the training and validation data to build \approach' predictive model. \approach then consumes that raw security event data generated from millions of machines that send back their activity reports. The collection and preprocessing module reconstructs the security events observed on a given machine $s_i$ as a sequence of events ordered by timestamps, in the format of $s_i=\{e^{(i)}_{1}, e^{(i)}_{2}, ..., e^{(i)}_{n}\}$. The output of the data collection and preprocessing module is $D = \{s_1, s_2, ..., s_m\}$ where $m$ denotes the number of machines. Finally, we build the training data $D_T$ and validation data $D_V$ from $D$ for the next stage, where $D_T \cap D_V = \emptyset$.

\noindent \textbf{Model training and validation.} The core of \approach consists of the training of a recurrent neural network with recurrent memory cells (\ding{183}.a, see Section~\ref{sec:lstm_ma_theory} for technical details about recurrent memory cells). 
Essentially, \approach specifies a probability distribution of $e_{w+1}$ possible events given historical observed events \{$e_{1}, ..., e_{w}$\}, where $w$ refers to the rollback window size, by applying an affine transformation to the hidden layer followed by a $softmax$,

\begin{equation}
Pr(e_{w+1} | e_{1:w}) = \frac{\exp(h^w \cdot p^j + q^j)}{\sum_{j' \in E} \exp(h^w \cdot p^{j'} + q^{j'}) }
\end{equation}

where $p^j$ is the $j$-th column of output embedding $P \in R^{m \times |V|}$ and $q^j$ is a bias term. Given the training data $D_T$, \approach' training objective is therefore to minimize the negative log-likelihood $\mathcal{L}$ of all the event sequences:

\begin{equation}
\mathcal{L} = - \sum_{t=1}^{|D_T|} Pr(e_{t} | e_{1:t-1} : \theta)
\end{equation}

\noindent We use the validation data $D_V$ to verify if the parameters $\theta$ identified during the training phase can achieve reasonable prediction performance (\ding{183}.b). It is important to note that $D_T$ and $D_V$ come from different machines so as to verify the general prediction capability of \approach on the endpoints that are not part of the training data.
\noindent \textbf{Security event prediction (\ding{184}).} Once the model is trained, \approach takes the historical events \{$e_{0}$, ..., $e_{i}$\} as the initial input (\ie a \emph{variable-length} input sequence inline with the real-world scenario) and predicts the probability distribution of $e_{{i+1}}$ given $E$ as $Pr[e_{{i+1}} | e_{0:i}] = \{e_1 : p_1, ~ e_2 : p_2, ~ ..., ~ e_{|E|} : p_{|E|}\}$. Our strategy is to sort $Pr[e_{{i+1}} | e_{0:i}]$  and choose the event with the \emph{maximum probabilistic score}. \approach then verifies with the actual event sequence whether $e_{{i+1}}$ is the correct prediction. In case of a wrong prediction, \approach updates its contextual information accordingly. Section~\ref{sec:case_studies} provides a detailed case study of the security event prediction phase in a real-world scenario.

\noindent \textbf{Prediction performance monitoring (\ding{185}).} Finally, in an effort to maintain the prediction accuracy as high as possible, the prediction performance monitor tracks and reports the evolution of different metrics, such as the Precision, Recall, and F1 of the current model. 
It is possible to elaborate such metrics on \approach' implementation in the wild as it is immediately possible to see whether \approach predicted the right event. If the predictions precision is dropping below a certain threshold, the system would automatically understand that is necessary to retrain the model.

\subsection{Recurrent Memory Array} \label{sec:lstm_ma_theory}

Long short-term memory (LSTM) and variants such as gated recurrent units (GRU) are the most popular recurrent neural network models for sequential tasks, such as in character-level language modeling~\cite{kim2016character}. One common approach to deal with complex sequential data is using a stacked RNNs architecture. Essentially, stacking RNNs creates a multi-layer feedforward network at each time-step, \ie the input to a layer being the output of the previous layer. In turn, stacking RNNs automatically creates different time scales at different levels, and therefore a temporal hierarchy~\cite{hermans2013training}. This approach has been proven practical and achieving good accuracy in various cases, such as log prediction~\cite{du2017deeplog}, binary function recognition~\cite{shin2015recognizing}, and function type recovery~\cite{chua2017neural}. Nevertheless, despite the proven success of stacked RNNs, one complication incurred by such strategy is the lack of generalization to new data, \eg stacking mechanisms chosen and tuned for current training data require vigorous evaluation and may not adapt well to the new data at run time~\cite{zhang2016architectural}. Therefore, rather than stacking multiple layers of RNNs, it would be ideal to build more complex memory structures inside a RNN cell to retain temporal memories while keeping a single layer RNN network to maintain computational efficiency when training. To achieve both goals, we propose to leverage the recurrent memory array by Rocki~\cite{rocki2016recurrent}; this is doable by modifying LSTM architectures, while it is not available on GRU architectures.

Following the notation in Rocki~\cite{rocki2016recurrent}, we can formally define the recurrent memory array as follows in Eq.~\ref{lstm_ma_1}.

\begin{eqnarray}\label{lstm_ma_1}
	\nonumber
	f_k^t & = & \sigma(W_{fk}x^t + U_{fk}h^{t-1} + b_{fk})\\ \nonumber
	i_k^t & = & \sigma(W_{ik}x^t + U_{ik}h^{t-1} + b_{ik})\\ \nonumber
	o_k^t & = & \sigma(W_{ok}x^t + U_{ok}h^{t-1} + b_{ok})\\ \nonumber
	\tilde{c}_k^t & = & tanh(W_{ck}x^t + U_{ck}h^{t-1} + b_{ck}) \\ \nonumber
	c_k^t  & = & f_k^t \odot c_k^{t-1} + i_k^t \odot \tilde{c}_k^t \\
	h^t & = & \sum_{k} o_k^t \odot tanh(c_k^t)
\end{eqnarray}

\noindent where $f$ denotes forget gates, $i$ denotes inputs, $o$ denotes outputs, $c$ denotes cell states, and $h$ denotes the hidden states. Here, $\odot$ represents element-wise multiplication. It is straightforward to notice that parameter $k$ directly controls the number of cell memory vectors, which enables the recurrent memory array to build an array-like structure similar to the structure of the cerebellar cortex~\cite{rocki2016recurrent}. 

To deal with noisy sequential input data (Section~\ref{sec:dataset}) as observed in the real world, we follow the \textbf{stochastic} design outlined in~\cite{rocki2016recurrent} by treating initial output gate activations as inputs to a $softmax$ output distribution, sampling from this distribution, and selecting the most likely memory cell to activate (see Eq.~\ref{lstm_ma_2}).

\begin{eqnarray}\label{lstm_ma_2}
	\nonumber
	p(i=k) & = & \frac{e^{o_k^t}}{\sum_k o_k^t} \\ 
	h^t & = & o_i^t \odot tanh(c_i^t)
\end{eqnarray}

Eq.~\ref{lstm_ma_2} identifies the probability of a memory cell $i$ to be activated and update $h^t$ accordingly using this cell while the rest of memory cells are deactivated. Hence, instead of summarizing all cell memory (see Eq.~\ref{lstm_ma_1}), only one output is used in this stochastic design that is resilient to noisy input. We refer interested readers to~\cite{rocki2016recurrent} for theoretical proofs and empirical comparison studies with the other state-of-the-art RNN architectures.

\section{Datasets} \label{sec:dataset}

\begin{figure}[t]
	\centering
	\includegraphics[width=0.8\linewidth]{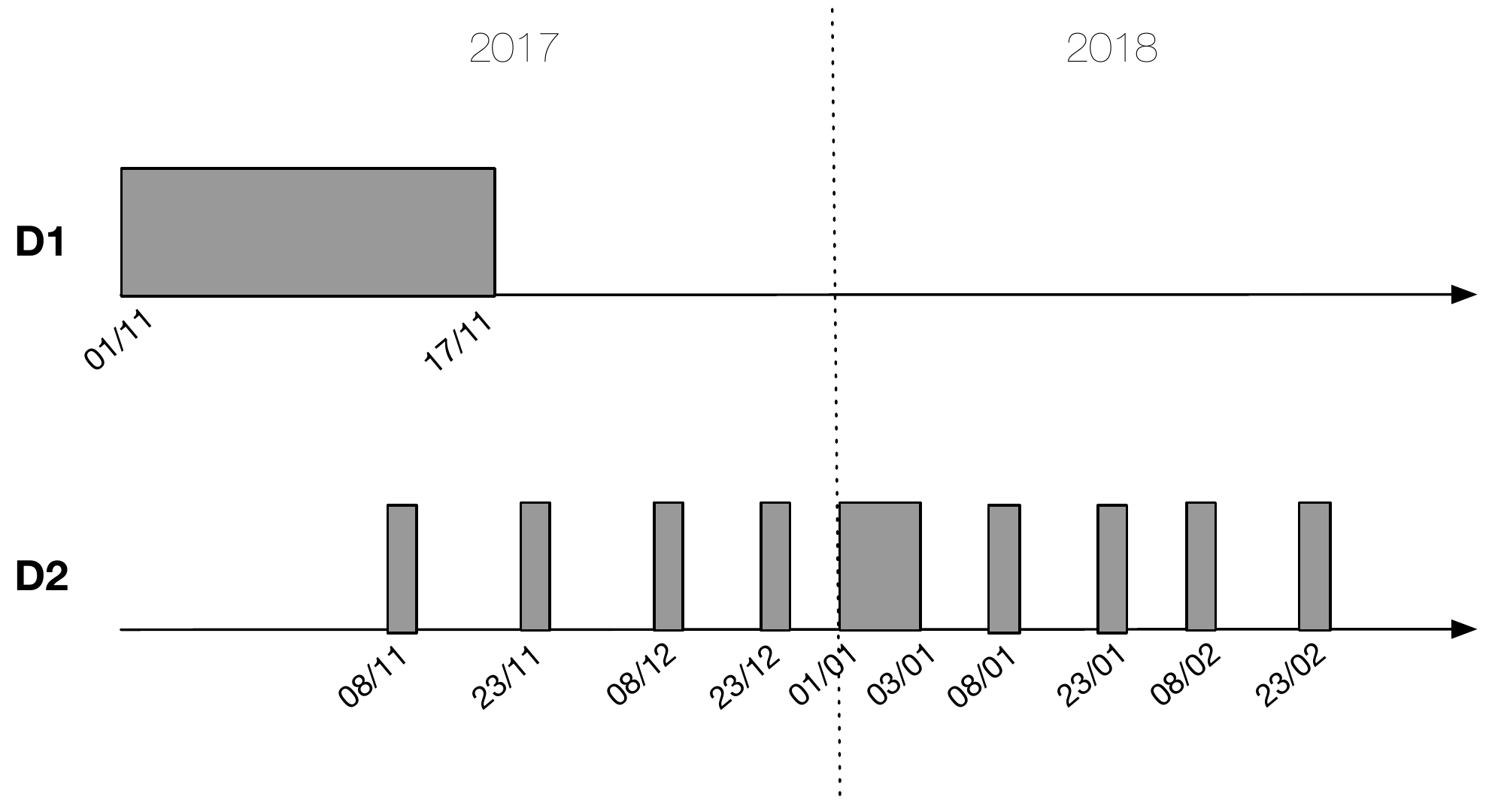}
	\caption{Summary of the security event datasets used in this paper. }
	\label{fig:ipsping_dataset}
  \vspace{-0.4cm}
\end{figure}

\approach is a generic system that can be used to predict security events on different protection systems. To evaluate its performance in this paper, we focus specifically on security event data collected from Symantec's intrusion prevention product. Symantec offers end users to explicitly opt in to its data sharing program to help improving its detection capabilities. To preserve the anonymity of users, client identifiers are anonymized and it is not possible to link the collected data back to the users that originated it. Meta-information associated with a security event is recorded when the product detects network-level or system-level activity that matches a pre-defined signature. From this data we extract the following information: anonymized machine ID, timestamp, security event ID, event description, system actions, and other information (e.g., file SHA2) if any. Note that we use the anonymized machine ID to reconstruct a series of security events detected in a given machine and discard it after the reconstruction process is done. 

To thoroughly investigate the effectiveness, stability and reusability of \approach, we collected 27 days of data, summarized in Figure \ref{fig:ipsping_dataset}. We compile two separate datasets. The first one, which we call $\mathbf{D1}$, spans a period of 17 days in November 2017 (1 November - 17 November), and is composed of over 2.2 billion security events. We use the first five days (1 November - 5 November) of $\mathbf{D1}$ to validate our approach and for a comparison study against three baseline methods (see Section \ref{sec:evaluation}). We later use the first seven days (1 November - 7 November) of $\mathbf{D1}$ to build models with varied length of training period, and study the stability of our approach by evaluating the prediction accuracy for the rest of the 10 days of data (8 November - 17 November) from $\mathbf{D1}$. We also compile another dataset, which we call $\mathbf{D2}$. This dataset is composed of 1.2 billion security events collected on the 8th and 23rd day of each month between November 2017 and February 2018, and the first three days in January 2018. $\mathbf{D2}$ is used to understand whether the system retains its accuracy even in a longer term scale: training sets based on $\mathbf{D1}$ are months older than part of the data in $\mathbf{D2}$. We use the first three days in January 2018 to build new models and compare them to the models built with data from $\mathbf{D1}$ (1 November - 7 November) and study their prediction performance with a focus on \approach' reusability (see Section \ref{sec:exp4}). On average, we collect 131 million security events from 740k machines per day, roughly 176 security events per machine per day. In total, the monitored machines generated 4,495 unique security events over the 27 day observation period. 

\noindent\textbf{Data Limitations.} It is important to note that the security event data is collected passively. That is, these security events are recorded only when corresponding signatures are triggered. Any events preemptively blocked by other security products cannot be observed. Additionally, any events that did not match the predefined signatures are also not observed. Hence the prediction model used in this paper can only predict the events observed by Symantec's intrusion prevention product.
We discuss more details on the limitations underlying the data used by \approach in Section~\ref{sec:limitations}.

\section{Evaluation} \label{sec:evaluation}
In this section we describe the experiments operated to evaluate \approach. 
We designed a number of experiments that allow us to answer the following research questions:

\begin{itemize}

  \item What is \approach performance in identifying the upcoming security event (Section~\ref{sec:overall_prediction}) and how does its performance compare to the baseline and state-of-the-art methods (Section~\ref{sec:comparison})?
  \item How do variations in the model's training period affect the performance (Section~\ref{sec:exp3})?
  \item Can we reuse a trained \approach model for a given period of time and when do we need to retrain the model (Section~\ref{sec:exp4})?
  \item What is the influence of the long-term memory of Recurrent Neural Network models to achieve security event prediction (Section~\ref{sec:seqlen})?

\end{itemize}

\subsection{Experimental Setup}
\label{sec:implementation}

\noindent \textbf{Implementation.} We implemented \approach in Python 2.7 and TensorFlow 1.4.1. Experimentally, we set the number of unrolling $w$ to 20, the training batch size to 128, the number of memory array $k$ (see Section~\ref{sec:lstm_ma_theory}) to 4 and the number of hidden LSTM Memory Array units to 128. We find these parameters offering the best prediction performance given our dataset. All experiments were conducted on a server with 4 TITAN X (Pascal) 12GB 1.5G GPUs with the CUDA 8.0 toolkit installed. All baseline methods are implemented in Python 2.7 and experimented on a server with a 2.67GHz Xeon CPU X5650 and 128GB of memory. 

\noindent \textbf{Evaluation setup.} To form a concrete evaluation setup, for both \approach and other baseline methods experiments, we split the input data and use 80\% for training, 10\% for validation, and 10\% for test. We strictly require that training, validation and test data to come from different machines so as to verify \approach' general prediction capability in the endpoints that are not part of the training data. Specifically, we train \approach for 100 epochs, validate model performance after \emph{every} epoch and select the model that provides the best performance on validation data. 

\noindent \textbf{Evaluation metrics.} We use the precision, recall, and F1 metrics to evaluate prediction results from the models. 
In our experimental setup, we calculate these metrics globally by counting the total true positives, false negatives and false positives. It is important to note that \approach accepts \emph{variable-length} security event sequences. We specially hold out the \emph{last} event as the prediction target $e_{tgt}$ for evaluation purposes. Section~\ref{sec:case_studies} showcases how \approach can be leveraged to accomplish step-by-step prediction with a single event as the initial input.

\subsection{Overall Prediction Results}
\label{sec:overall_prediction}

\begin{figure}[t]
	\centering
	\includegraphics[width=0.7\linewidth]{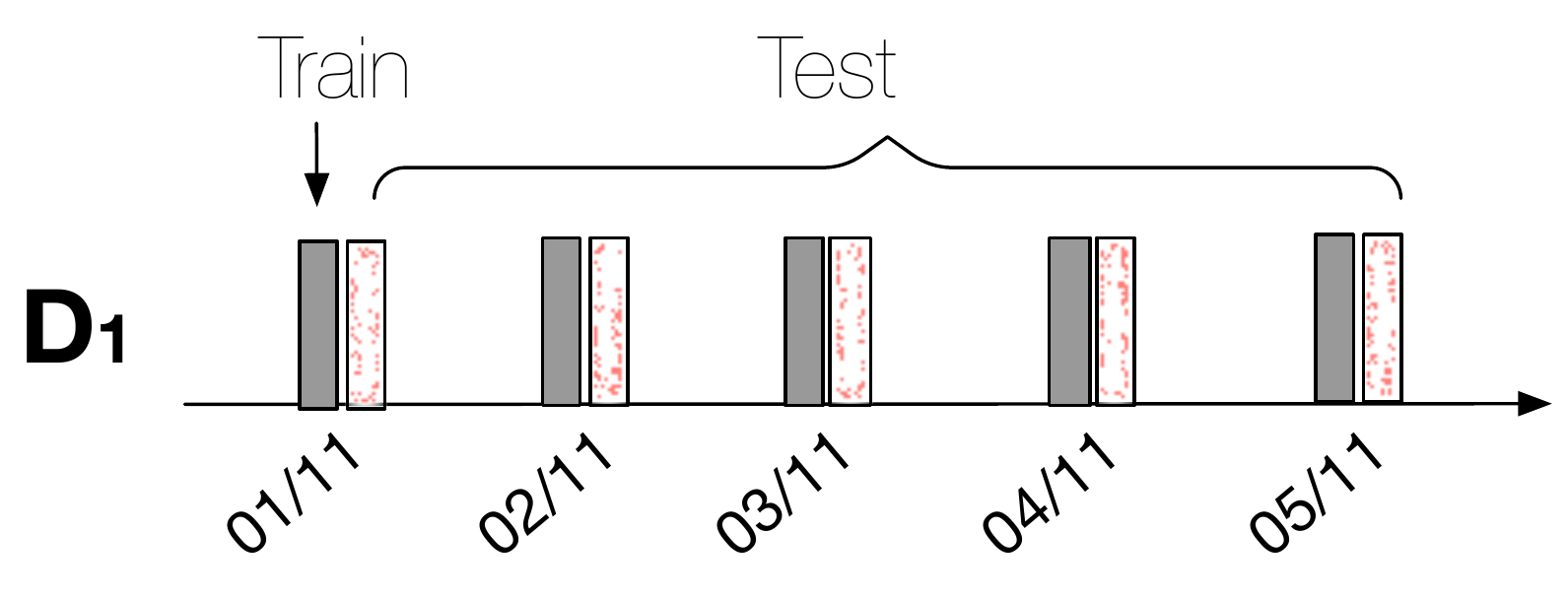}
	\caption{Experimental setup for \approach' prediction evaluation (Section~\ref{sec:overall_prediction}) and comparison study with baseline methods (Section~\ref{sec:comparison}).}
	\label{fig:exp1_evaluation}
	\vspace{-0.5cm}
\end{figure}

In this section we evaluate the performance of our security event forecast model in predicting the exact upcoming event. This is a challenging task that a predictive system for security events aims at resolving due to the fact that there are 4,495 security events as possible candidates in our dataset (see Section~\ref{sec:dataset}) and an exact event should be correctly predicted. 

\begin{figure*}
     \centering
    \begin{subfigure}[t]{0.3\textwidth}
        \raisebox{-\height}{\includegraphics[width=\textwidth]{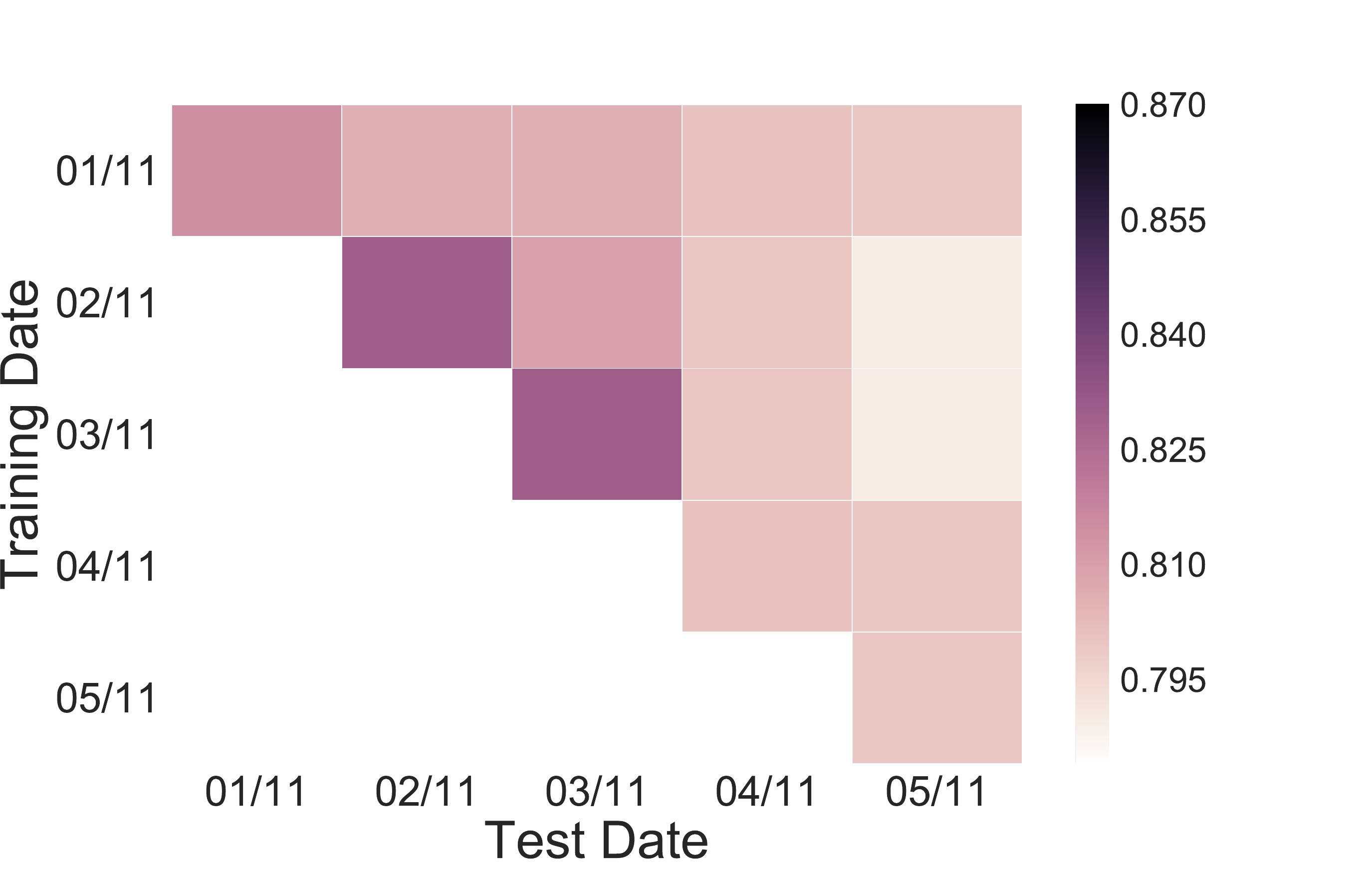}}
        \caption{Precision}
    \end{subfigure}
    \hfill
    \begin{subfigure}[t]{0.3\textwidth}
        \raisebox{-\height}{\includegraphics[width=\textwidth]{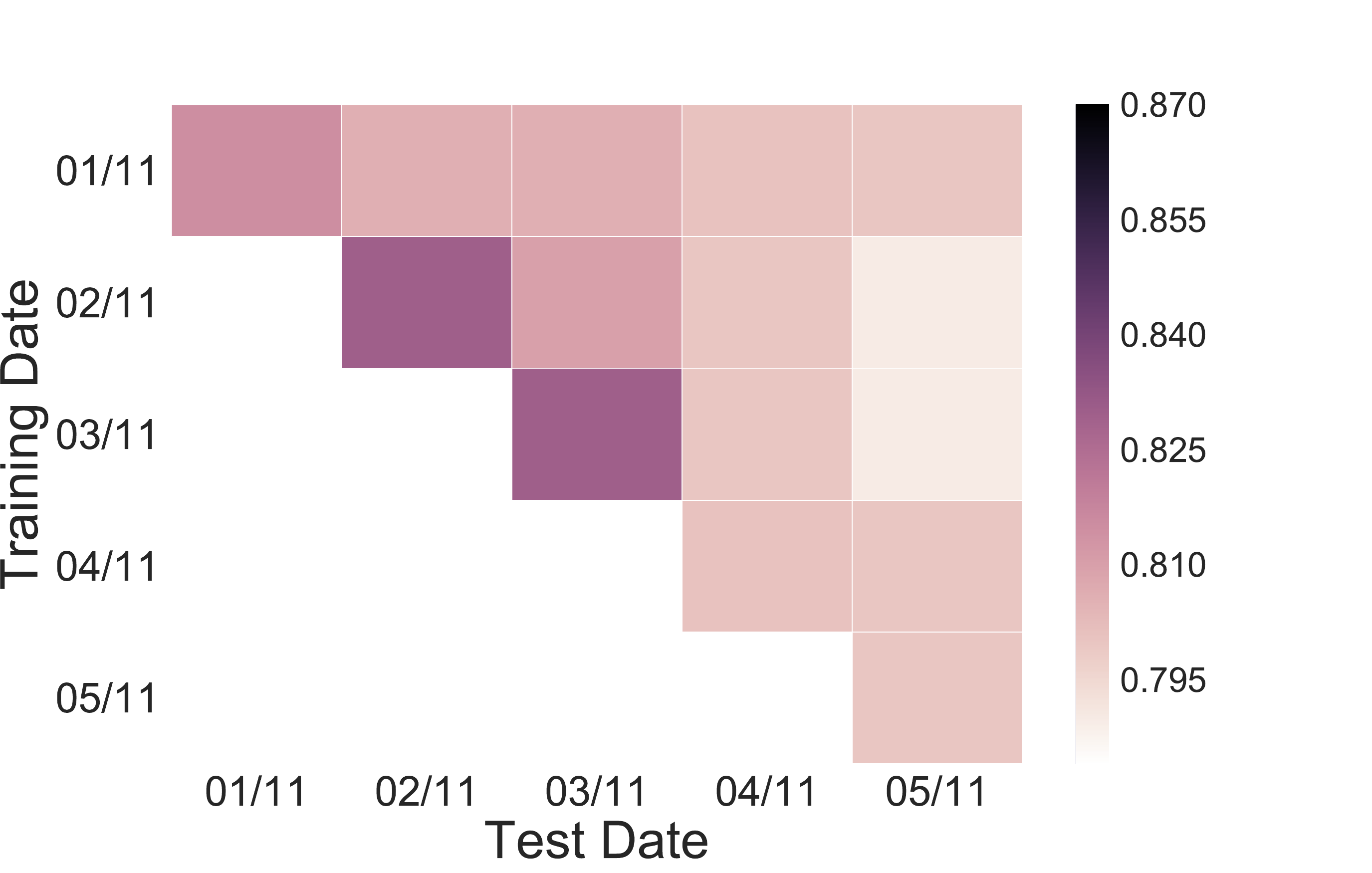}}
        \caption{Recall}
    \end{subfigure}
    \hfill
    \begin{subfigure}[t]{0.3\textwidth}
        \raisebox{-\height}{\includegraphics[width=\textwidth]{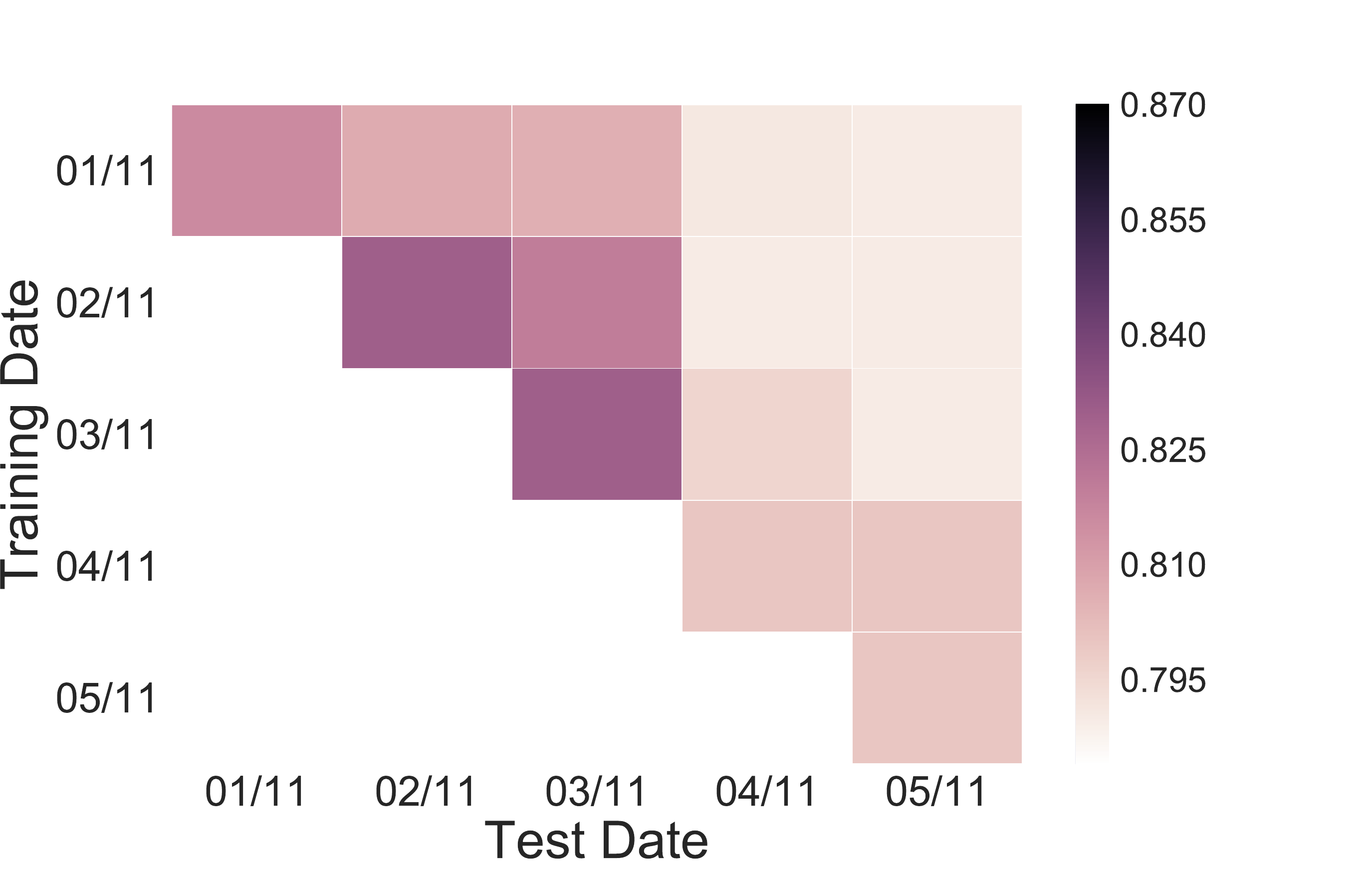}}
    \caption{F1-Measure} 
    \end{subfigure}
    \caption{Precision, Recall, and F1-Measure of overall \approach's performance. \approach is trained using one day of data and evaluated on both the same day and the following days until 5 November 2017.}
		\label{fig:NextDays}
		\vspace{-0.4cm}
\end{figure*}

\noindent \textbf{Experiment setup}. We use the experimental setup as illustrated in Figure~\ref{fig:exp1_evaluation} for \approach' performance evaluation. From $\mathbf{D1}$, we train our predictive model using \textbf{one day} of data and evaluate \approach on both the same day and the following days until 5 November 2017. For example, we train \approach using data from 2 November and evaluate its prediction performance from 2 November to 5 November.

\noindent\textbf{Experiment results.} Following our general experimental evaluation setup, we randomly select $14,396$ machines from the first days of November that are not part of the data used in the training set of the initial model. We focus on predicting the last event occurring on a machine given the sequence of previously-observed events. 
As shown in Figure~\ref{fig:NextDays}, \approach is able to achieve over 80\% precision, recall, and F1-measure in predicting the exact upcoming security event when evaluating on the same day test data. 
Figure~\ref{fig:NextDays} shows that, when training on one day, and testing on the same day and the following ones, the values of the Precision, Recall and F1 do not decrease dramatically. When it does, it decreases, in the worse case, of less than $0.05$. The Figure also shows that Precision (Figure~\ref{fig:NextDays}a) and Recall (Figure~\ref{fig:NextDays}b) are well balanced and have very similar values and exactly the same scale (from $0.87$ to $0.795$). This result shows that \approach can offer good prediction results. 
There is no security event dominating in our dataset, which may lead to biased but better prediction performance. The top 3 events in our training data are: (i) Microsoft SMB MS17-010 Disclosure Attempt (19.8\%), (ii) SMB Double Pulsar Ping (16.4\%), and (iii) Unimplemented Trans2 Subcommand (16.1\%). The top 3 events in our test data are ranked as follows: (i) Microsoft SMB MS17-010 Disclosure Attempt (9.85\%), (ii) HTTP PE Download (6.3\%), and (iii) DNS Lookup Failures (3.5\%). Interestingly, the dominant events in training and test are different, which makes \approach' prediction results even stronger.

Over the days, we observe a trend that the prediction performance of \approach drops slightly in terms of all three evaluation metrics. Take the model trained on 2 November for example, its prediction precision drops by 4\% from 0.83 to 0.79. In Section~\ref{sec:exp3} we study if variations (\eg a longer training period) in the model's training data would offer better performance and how stable the trained \approach performs over consecutive days. 
Note that `micro'-averaging in a
multi-class setting produces equal Precision, Recall and F1-Measure. For the rest of the evaluation process, we therefore use precision as the main evaluation metric. 

\subsection{Comparison Study}
\label{sec:comparison}

\begin{table}[]
\centering
\footnotesize

\begin{tabular}{|c|c|c|c|c|c|}
\hline
                                                       & \multicolumn{5}{c|}{\textbf{Test Date (Evaluation Metric - Precision)}}             \\ \hline
\textbf{Method}                                                 & \textbf{01/Nov} & \textbf{02/Nov} & \textbf{03/Nov} & \textbf{04/Nov} & \textbf{05/Nov} \\ \hline
Spectral                                               &    0.05    &   0.031     &   0.023     &  0.013      & 0.02       \\ \hline
\begin{tabular}[c]{@{}c@{}}Markov Chain\end{tabular} &   0.62    &  0.56      &  0.56      &  0.53      &   0.52     \\ \hline
3-gram                                                 &     0.67   & 0.54       &  0.61      &  0.592      & 0.601       \\ \hline
\approach                                              &   \textbf{0.83}     & \textbf{0.82}      &  \textbf{0.83}      &  \textbf{0.82}      &  \textbf{0.81}      \\ \hline
\end{tabular}

\caption{Prediction precision comparison study: \approach vs. baseline approaches.}
\label{tab:sameday}
\vspace{-0.6cm}
\end{table}

In this section we aim at studying whether the higher complexity of Recurrent Neural Networks is required for the task of predicting security events, or whether simpler baseline methods would be enough for the task at hand.
For comparison purposes, we implemented \emph{first-order} Markov Chain~\cite{norris1998markov} and 3-gram model~\cite{brown1992class} (equivalent to the second order Markov Chain model) in Python 2.7.1. Note that it is natural to consider a higher order (\eg n-order where $n > 2$) Markov Chain model for security event prediction, however, due to the exponential states issue associated with high order Markov Chain models, it is computationally costly to build such a high order model for 4,495 events. Finally we use the sp2learning\footnote{https://pypi.org/project/Sp2Learning/1.1.1/} package to build a spectral learning model~\cite{kamvar2003spectral} for sequence prediction as the third baseline prediction model. These three methods are often used to model sequences of elements in several fields and, being simpler than our RNN models and widely used in sequence prediction, they are relevantly good baselines to compare \approach with.

\noindent \textbf{Experiment setup.} The comparison study uses daily data (1 November - 5 November) from $\mathbf{D1}$. To evaluate \approach in this case, all training, validation, and test data come from the same day. 

\noindent \textbf{Comparison study results.} Table~\ref{tab:sameday} shows the precision of \approach compared to simpler systems. 
Table~\ref{tab:sameday} shows that \approach outperforms the baseline methods but also that 3-grams perform better than Markov Chains, and Markov Chains perform better than the spectral learning method. This particular order shows the importance of sequence memory as the system that performs best among the baselines is the 3-grams. However, 3-grams are less effective than \approach. This is due to two of the main characteristics of neural networks: the capacity of filtering noise and the longer term memory. As Table~\ref{tab:sameday} shows, \approach has precision values higher than 0.8 in all the five days of tests showing a very good level of reliability. In Section~\ref{sec:seqlen} we show that the long-term memory that is an important feature of RNNs plays a key role in correctly predicting security events. Note that we didn't report a comparison of the computation time among the methods due to the fact that \approach leverages GPUs to train RNN models and the baseline methods rely on traditional CPUs, and therefore \approach is in general much quicker to run. For example, our 3-gram implementation took over 10 days for training, yet \approach requires only $\sim$10 minutes per epoch using GPUs. 

\subsection{Influence of Training Period Length}
\label{sec:exp3}

\begin{figure}[t]
	\centering
	\includegraphics[width=0.85\linewidth]{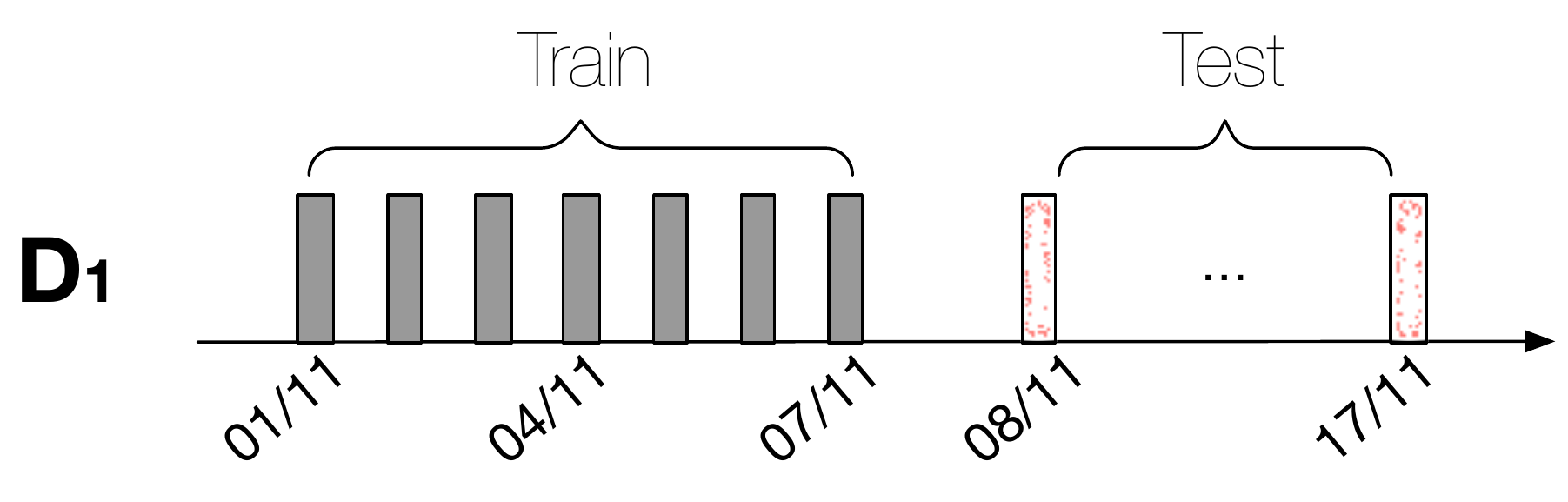}
	\caption{Experimental setup for multiple day evaluation of \approach (Section~\ref{sec:exp3}).}
	\label{fig:exp3_evaluation}
	\vspace{-0.4cm}
\end{figure}

In this section we look at whether training \approach on longer periods of time achieves better prediction performance.

\noindent \textbf{Experiment setup.}  We use the experimental setup as illustrated in Figure~\ref{fig:exp3_evaluation} for \approach' performance evaluation. From $\mathbf{D1}$, we train our predictive model using \textbf{one day} of data and evaluate on the test data from 8 November to 17 November. For example, we train \approach using data from 2 November and evaluate its prediction performance on test data from 8 November to 17 November. To evaluate if a longer training period can offer better prediction performance, we also train our predictive model using \textbf{one week} of data (from 1 November to 7 November) and evaluate its performance in the aforementioned period.

\noindent\textbf{Experiment results.} In this experiment we evaluate the performance of our security event forecast model in predicting the exact upcoming event several days after the initial model was trained. The goal is to determine how well our predictive model ages in the short term and to make sure that it remains effective in predicting security events without the need to re-tune it after this period of time.

The question that this experiment is trying to answer is whether there is a difference in training the models over longer periods of time, such as one week, rather than one day. Table~\ref{tab:NovTests} provides some insights into this question. First, we used the first \textbf{five} days of November on their own to build five models. Second, we built one single model from the first \textbf{seven days} of the same month. We then tested the six different models (five based on one day of data and one based on one week of data) on \textbf{ten days} of data from 8 November to 17 November. \emph{Overall, the training over one week of data produces similar results as those obtained using training over only one day of data.} On average, \approach trained with one week data can achieve a precision score of 0.819, which is 0.3\% higher than that of the models trained with one day data.

These results demonstrate that \approach can offer good accuracy with stable performance over time since the standard deviation of precision scores over the measurement period of 10 days is small ($\sim$0.02). However, on 8 November and on 16 November the results are slightly different, exhibiting a higher accuracy for the week-long trained model. While in the first case (8 November) it is probably due to the proximity of the test day to the training week, the second case (16 November) appears to be an outlier. 
We further observe that the time proximity of the training and test data appears to have a positive impact on the prediction accuracy. Indeed, we can see that the model trained over one day of data is as efficient as the one trained over one week of data when tested on alerts generated only a few days later, probably due to the similarity among attack behaviors observed within a few days. The week-long trained model appears to be more efficient in the presence of deviating, or outlying attack behaviors in the test phase. This can easily be explained by the fact that the more data is used to build a model the more complete the model is. Hence it can better deal with rare events or deviating attack behaviors. 

One of the reasons why \approach' prediction precision might suddenly decrease is if the set of alerts significantly changes from a day to another, for example because a new vulnerability starts being exploited in the wild, a system patch fixes an existing one, or a major version of a popular software gets released. For this reason, in our architecture discussed in Section~\ref{sec:workflow} we included a component that monitors the performance of \approach and can trigger a re-training of the system if it is deemed necessary. 
In the experiment discussed in Table~\ref{tab:NovTests}, for example, the precision performance on 16 November drops by 6.9\% on average from 8 November. This could indicate to the operator that something significant changed in the monitored systems and that \approach needs to be retrained. 
As we will show in Section~\ref{sec:perform}, this can be done in batch and it takes well less than a day to complete.

\begin{table*}[]
\centering
\resizebox{0.7\linewidth}{!} {
\begin{tabular}{|c|c|c|c|c|c|c|c|c|c|c|}
\hline
                                                              & \multicolumn{10}{c|}{\textbf{Test Date (Evaluation Metric: Precision)}}             \\ \hline
\begin{tabular}[c]{@{}c@{}}\textbf{Model}\\ \textbf{Training Date}\end{tabular} & \textbf{08/Nov} & \textbf{09/Nov }&\textbf{ 10/Nov} & \textbf{11/Nov} & \textbf{12/Nov} & \textbf{13/Nov} & \textbf{14/Nov} & \textbf{15/Nov} & \textbf{16/Nov} & \textbf{17/Nov} \\ \hline
01/Nov                                                        &   0.815     &  0.823      &  0.822      &  0.794      & 0.789   & 0.814 & 0.817 & 0.816 & 0.746 & 0.774    \\ \hline
02/Nov                                                        &   0.821     &  0.827      &  0.826      &  0.801      & 0.792  & 0.82 & 0.819 & 0.82  & 0.76  & 0.79    \\ \hline
03/Nov                                                        &   0.822     &  0.827      &  0.826     &    0.80    &   0.794   & 0.82 & 0.82 & 0.817 & 0.742 & 0.769  \\ \hline
04/Nov                                                        &   0.820     &  0.828      &  0.827      &   0.797     & 0.797    & 0.822 & 0.823 & 0.82 & 0.75 & 0.77   \\ \hline
05/Nov                                                        &   0.817     &  0.825      &  0.823      &   0.791     &  0.791   & 0.818 & 0.815 & 0.815 & 0.747 & 0.775   \\ \hline
01/Nov - 07/Nov                                               &   0.836     &  0.83      &  0.823      &   0.82     &  0.801  & 0.815 & 0.816 & 0.812 &  0.783 & 0.773      \\ \hline
\end{tabular}
}
\caption{Evaluation of \approach' prediction precision between 8th November and 17th November.}
\label{tab:NovTests}
\vspace{-0.7cm}
\end{table*}

\subsection{Stability Over Time}
\label{sec:exp4}

\begin{figure}[t]
	\centering
	\includegraphics[width=0.9\linewidth]{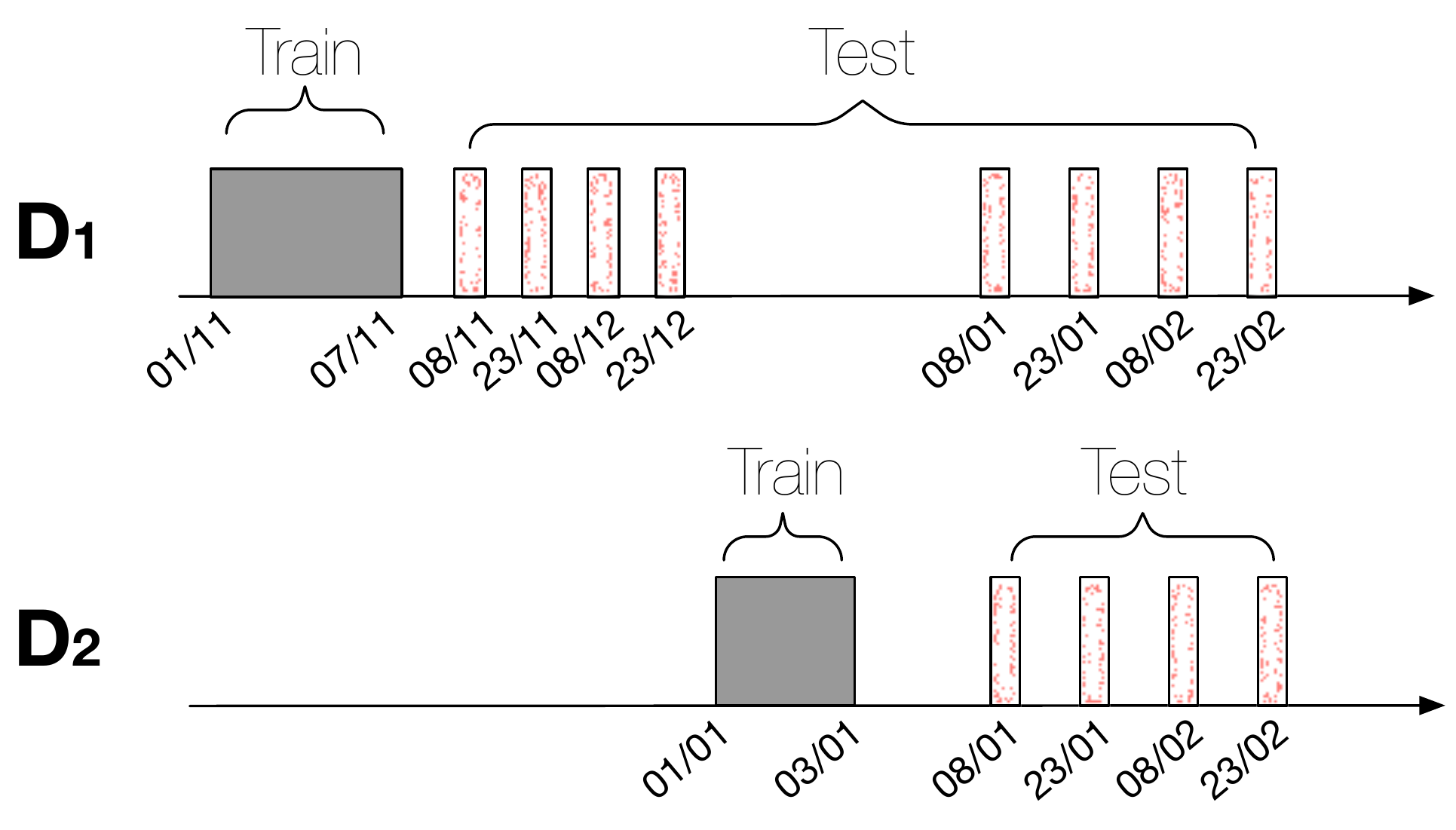}
	\caption{Experimental setup for \approach reliability evaluation (Section~\ref{sec:exp4}).}
	\label{fig:exp4_evaluation}
	\vspace{-0.4cm}
\end{figure}

In this Section we evaluate \approach' prediction accuracy when the training data is several months older than the test data. Our goal is to evaluate the reliability of the model in case there is no retraining for several months. 
As we discussed, \approach is able to detect when it needs to be retrained, however this operation does not come for free and therefore it is desirable to minimize it as much as possible.

\noindent \textbf{Experiment setup.} The experimental setup is illustrated in Figure~\ref{fig:exp4_evaluation}. We train our predictive model using both \textbf{one day} of data (from 1 November to 5 November respectively) and \textbf{one week} of data (from 1 November to 7 November) from $\mathbf{D1}$. Additionally for comparison purposes, we train three more predictive models using \textbf{one day} of data (from 1 January to 3 January respectively) from $\mathbf{D2}$. The test data consists of two days per month (on the 8th and the 23rd) so as to obtain a representative dataset from November 2017 until February 2018.

\noindent \textbf{Experiment results.} Table~\ref{tab:MonthsTests} shows the results obtained using the same training sets as in the previous Section augmented with three days in January, \ie one day-long model for each of the first five days of November 2017, one week-long model for the first seven days of the same month and one day-long model for each of the first three days of January 2018.  The prediction precision results presented in Table~\ref{tab:MonthsTests} show consistency through the different training sets and a good level of stability, as the performance does not decrease dramatically over time. Moreover, the week-long training set does not show increased accuracy compared to the day-long ones. These new results thus confirm those from Section~\ref{sec:overall_prediction} and show that \emph{(i) the model quickly converges towards high accuracy with only one or a few days of training data, and (ii) the model ages very well even months after it was built}.

\noindent\emph{December discontinuity.} Table~\ref{tab:MonthsTests} shows a particular behavior between 8 December and 23 December: \approach' precision increases. We would normally expect the system's precision to slightly decrease over time, possibly following a pattern, while in this case the precision increases. To investigate this phenomenon, we looked for potential differences in the raw data and noticed that the test data collected after 8 December exhibits a significant deviation with respect to one specific security event ID: the presence of one of the top three recorded alarms decreased by an order of magnitude, having a comparable number of occurrences to alerts occupying the 4th to 10th position. The alarm is related to DoublePulsar, a vulnerability disclosed in the first half of 2017. Such change may be due to different reasons. The most probable reason, however, could be the installation of patches: Microsoft releases monthly updates for Windows every 2nd Tuesday of the month (\eg 12 December 2017) and many software- and hardware-related companies release patches immediately following Microsoft's. Finally, a small change to the IPS signatures or to the attack modus operandi can heavily impact the hit rate of a given alarm.

\noindent\emph{Comparison study.} To further investigate this December discontinuity phenomenon we decided to assess the impact of the training data on the model accuracy. To this end, we considered the training sets from data collected on the first three days of January and tested on the January and February dates (bottom part of Table~\ref{tab:MonthsTests}). We can see that \approach trained in January performs slightly better than when trained in November.
These results show that the results by \approach remain reliable even months after the system was trained. Nevertheless, in the case of a sudden decrease in precision due to an adverse change in the data (e.g., the emergence of a new attack), \approach would be able to detect this and prompt a retraining, as discussed in Section~\ref{sec:workflow}.

\begin{table*}
\centering
\resizebox{0.6\linewidth}{!} {
\begin{tabular}{|c|c|c|c|c|c|c|c|c|}
\hline
\multirow{3}{*}{\begin{tabular}[c]{@{}c@{}}\textbf{Model}\\ \textbf{Training Date(s)}\end{tabular}} & \multicolumn{8}{c|}{\textbf{Test Date (Evaluation Metric: Precision)}}          \\ \cline{2-9} 
                                                                               & \multicolumn{4}{c|}{\textbf{2017}}     & \multicolumn{4}{c|}{\textbf{2018}}     \\ \cline{2-9} 
                                                                               & \textbf{08/Nov} & \textbf{23/Nov }& \textbf{08/Dec} & \textbf{23/Dec} & \textbf{08/Jan} & \textbf{23/Jan} & \textbf{08/Feb} & \textbf{23/Feb} \\ \hline
01/Nov                                                                          & 0.815& 0.785 & 0.832 & 0.899 & 0.899 & 0.921 & 0.93 & 0.921      \\ \hline
02/Nov                                                                          & 0.821& 0.8 & 0.835 & 0.895 & 0.896 & 0.921 & 0.931 & 0.918       \\ \hline
03/Nov                                                                          & 0.822& 0.782 & 0.835 & 0.898 & 0.899 & 0.923 & 0.93 & 0.922      \\ \hline
04/Nov                                                                          & 0.820& 0.793 & 0.834 & 0.901 & 0.898 & 0.922 & 0.929 & 0.921       \\ \hline
05/Nov                                                                          &0.817 & 0.79 & 0.833 & 0.9 & 0.898 & 0.921 & 0.929 & 0.92       \\ \hline
01/Nov-07/Nov                                                                    & 0.836& 0.788 & 0.829 & 0.895 & 0.892 & 0.917 & 0.925 & 0.915      \\ \hline \hline
01/Jan                                                                    & - & - & -  & - & 0.905 & 0.927 & 0.931 & 0.926      \\ \hline 
02/Jan                                                                    & - & - & -  & - & 0.908 & 0.926 & 0.930 & 0.924      \\ \hline 
03/Jan                                                                    & - & - & - & - & 0.914 & 0.933 & 0.935 & 0.929      \\ \hline 
\end{tabular}
}
\caption{Evaluation of \approach's prediction precision on every 8th and 23rd of each month.}
\label{tab:MonthsTests}
\vspace{-0.7cm}
\end{table*}

\subsection{Sequence Length Evaluation}
\label{sec:seqlen}

In Section~\ref{sec:comparison} we showed that \approach outperforms simpler systems that do not take advantage of long-term memory in the same way as the RNN model used by our approach. 
In general, understanding how Deep Learning models work is challenging, and they are often treated as black boxes.
To make matters more complex, RNNs do not only rely on long-term memory, but also on short-term memory, in particular to filter out noise.

In this section we aim at understanding whether long-term memory is more influential in making decision than short-term memory or vice versa.
With relying on short-term memory we mean a system that relies on a few elements of the sequence to make its decision, that is, the ones closest to the element that the system is trying to guess.
With relying on long-term memory we mean when the system uses the whole sequence or a large part of it to take its decision on what the next security event could be.
Intuitively, if short-term memory was predominant, we would not expect the performance of \approach to increase with the number of observed events.

\begin{figure}[h]
    \centering
    \begin{subfigure}[b]{0.45\textwidth}
        \centering
        \includegraphics[width=\textwidth]{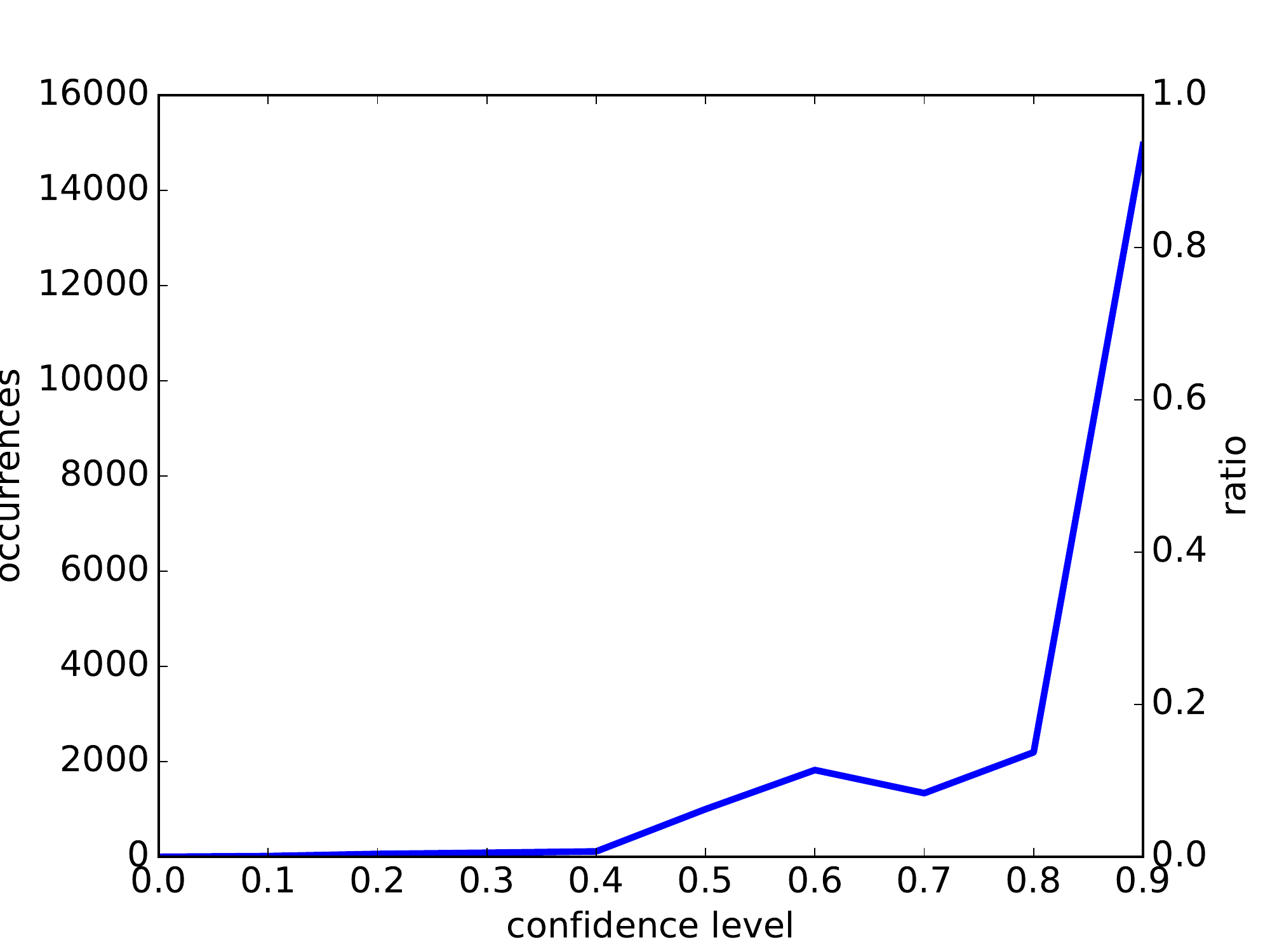}
        \caption{}
        \label{fig:SuccOcc}
    \end{subfigure}
    \hfill
    \begin{subfigure}[b]{0.45\textwidth}
        \centering
        \includegraphics[width=\textwidth]{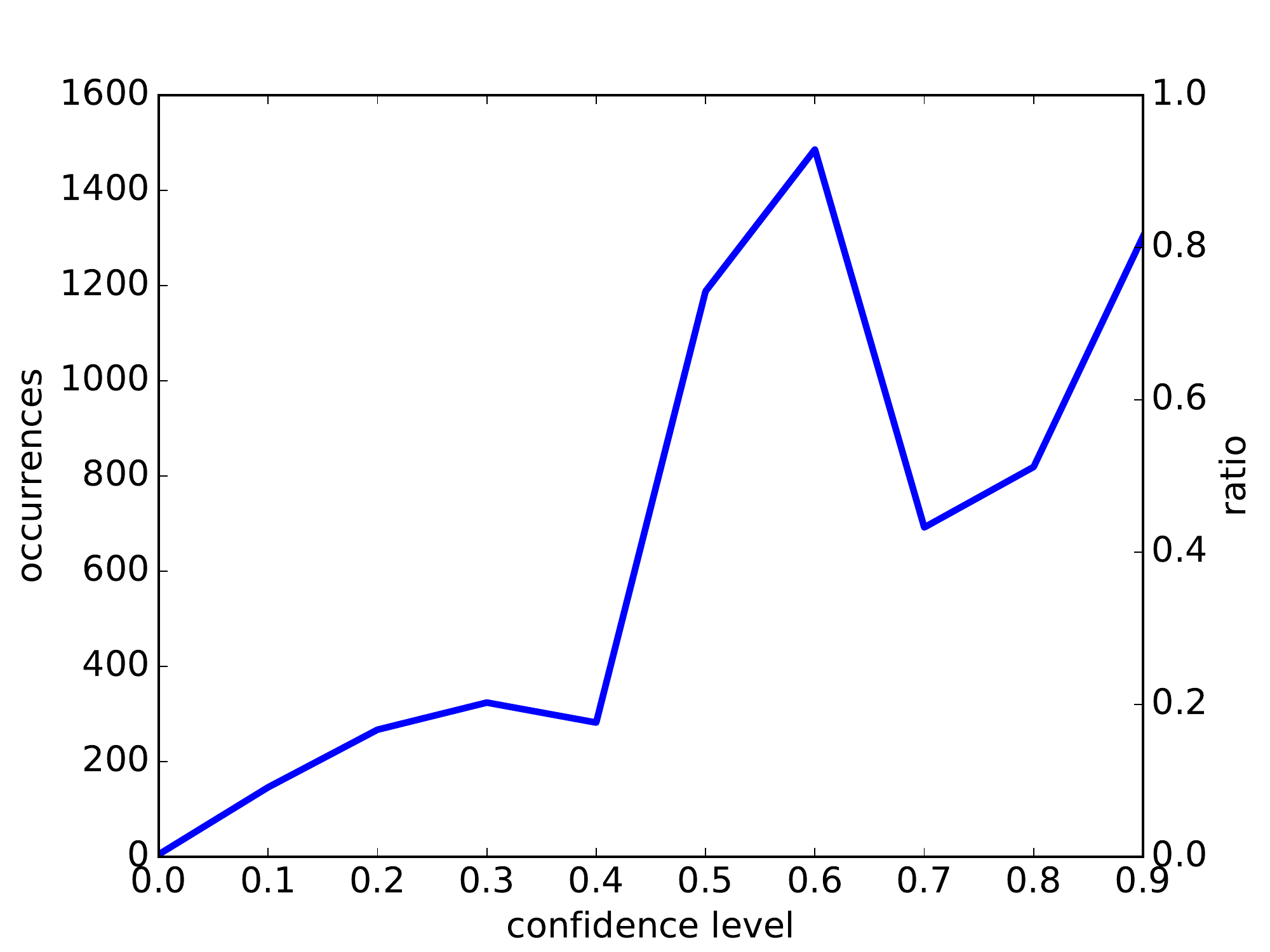}
        \caption{}
        \label{fig:FailOcc}
    \end{subfigure}
	\caption{Quantity of successfully and unsuccessfully guessed events. The Y axis on the left of each graph is the occurrence of successes/failures with at least the probability indicated on the X axis according to the system. The Y axis on the right is the ratio between the value on the other Y axis and the total of successes/failures.}
	
\end{figure}

As looking into the Neural Network weights may be a complicated way to understand which type of memory is more important for the model, we decided to focus on the occurrences of successfully and unsuccessfully guessed events.
Every event guessed by \approach has a probability (confidence score) associated to it. First, we look at the distribution of the confidence scores among successfully guessed events (Figure~\ref{fig:SuccOcc}) and unsuccessfully guessed ones (Figure~\ref{fig:FailOcc}).
As it can be seen, both types of events present a very skewed distribution in their confidence scores, with a negligible number of events being predicted with a probability of less than 0.5.

\begin{figure}[h]
    \centering
    \begin{subfigure}[b]{0.45\textwidth}
        \centering
        \includegraphics[width=\textwidth]{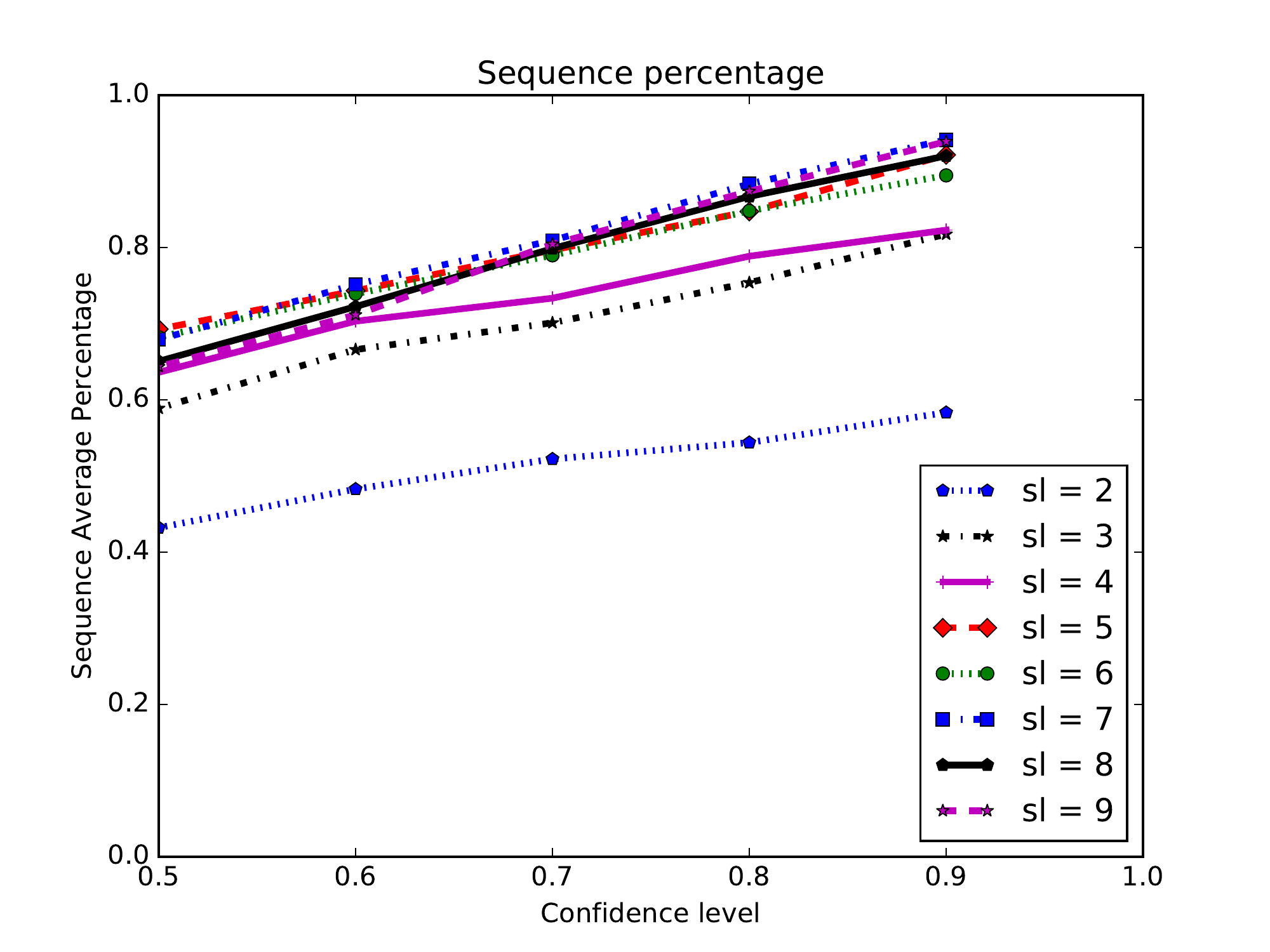}
        \caption{}
        \label{fig:SuccGuess}
    \end{subfigure}
    \hfill
    \begin{subfigure}[b]{0.45\textwidth}
        \centering
        \includegraphics[width=\textwidth]{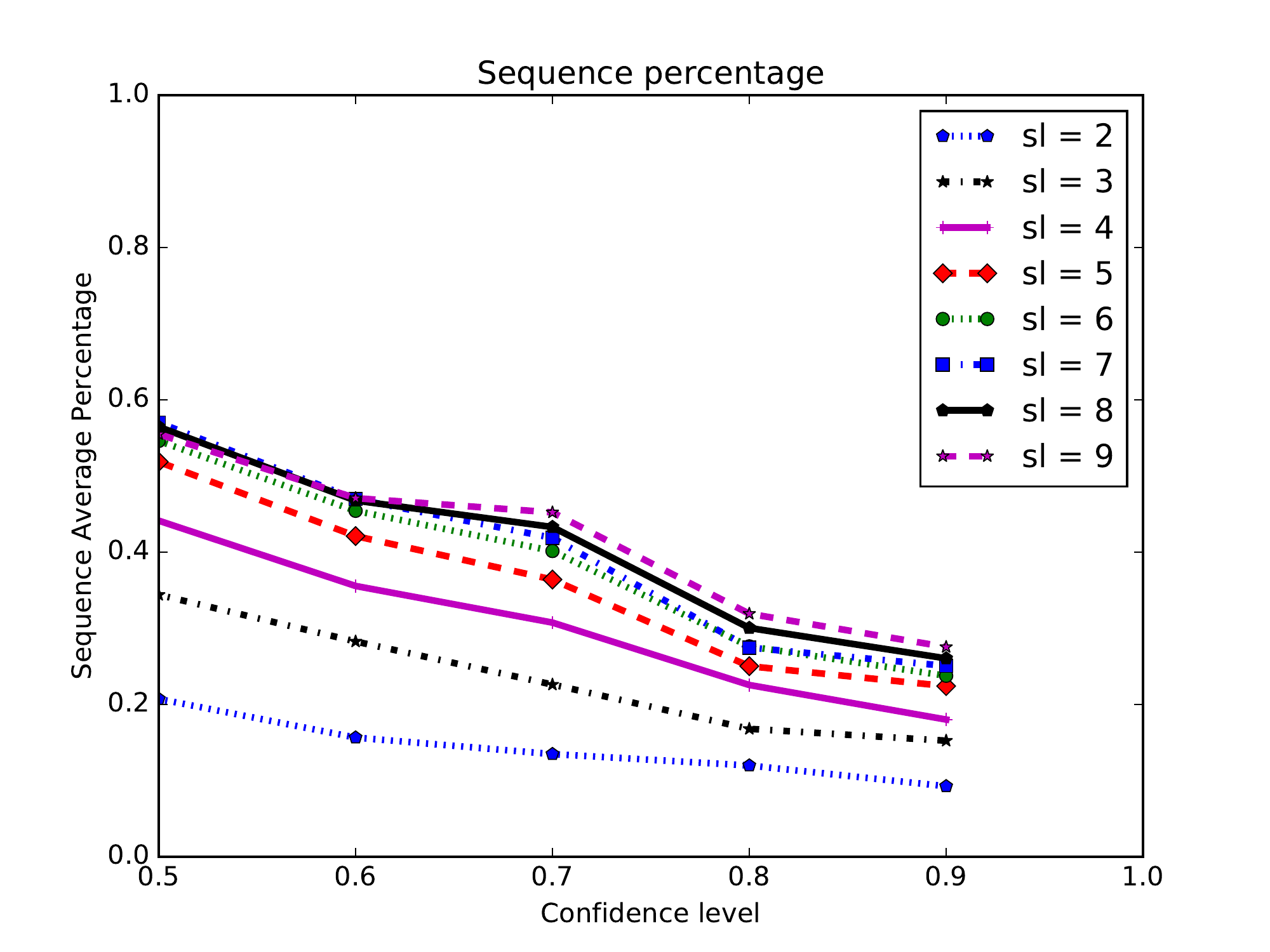}
        \caption{}
        \label{fig:FailGuess}
    \end{subfigure}
	\caption{The plots show the percentage of the sequences correctly guessed (a) or failed to guess (b) with respect to sequences that share all the events but the last. On the X axis, as for Figures~\ref{fig:SuccOcc} and~\ref{fig:FailOcc} there is the confidence level of the sequences used by the system.
	Figures show that sequences of at least 5 events (sl >=5) are quite unique, therefore long term memory is a crucial factor in the system accuracy.} 

\end{figure}

To better understand if \approach' results are mainly due to the use of long-term rather than short-term memory, we checked how unique the sequences on which \approach makes its decisions are. These quantitative results can hint at which kind of approach is used by the algorithm. We try to evaluate the occurrences of the sequence in which the system tried to guess the last event compared to all those that differ from it for the last event (the one that \approach tried to guess). This analysis is carried out for sequences of length $i+1, (i=2,...,9)$ where $i$ represents the number of events before the last we take into consideration. For all the successful/unsuccessful sequences we calculate the ratio between the times in which we had those $i$ events and \approach predicted the last event correctly and the times in which we had the same $i$ events followed by any event (included the right one). According to the probability value of the guessed event, we calculate the average probability.

Figure~\ref{fig:SuccGuess} shows the data for all the sequences for which the last event has been correctly guessed by \approach. Note that the X axis starts at 0.5 because, as Figures~\ref{fig:SuccOcc} and~\ref{fig:FailOcc} showed, the number of predictions with a lower confidence is very low. The values for $i$ (sequence length) less than 5 show that the system's prediction is not very confident. Longer sequences ($i$ greater than 5), instead, are more unique and often correctly predict the last event. 
The opposite happens when we evaluate the sequences involving events not guessed correctly by our system (Figure~\ref{fig:FailGuess}). In fact, the left part of the graph presents sequences where the wrongly guessed event was rarely the one following the $i$ previous events. This may mean that in those cases there are sequences that differ only for the last event and a few events are quite frequent.

\noindent\textbf{Takeaways.} Long sequences including the guessed event are more frequent when we analyze the successful guesses. This situation is more common than the unsuccessful guesses as the system reaches high accuracy. Therefore, according to the graphs the system seem to rely more on long-term memory than short-term memory.

\subsection{\approach Runtime Performance}
\label{sec:perform}

We now discuss the specific characteristics of the system and its runtime performance.

The training phase is the longest one: building a \approach model is a long process that can be performed offline. \approach takes around 10 hours to retrain the model. Considered the stability of the model, which as shown in Section~\ref{sec:exp4} can be effective for long periods of time, rebuilding the model does not have to be done every day. We also showed that it is possible to identify when the system needs retraining because of  a discontinuity in the distribution of events (see Section~\ref{sec:exp3}). Once trained, \approach takes 25ms to 80ms to predict the upcoming event using the variable-length security events in a given system.

\approach' predictive model trained using one day of data is about 31MB. It can be easily pushed to the endpoints with limited network footprint.
Note that with the advance of deep learning libraries, especially the recent development of TensorFlow, it is feasible for \approach to be deployed not only in traditional endpoints (\eg PCs) but in mobile and embedded devices as well. This is another aspect that exemplifies the general applicability of \approach.

\section{Case Studies}\label{sec:case_studies}

In this section we describe a set of case studies showcasing the capabilities of \approach in different real-world scenarios.
We first show how \approach can be used to detect a coordinated multi-step attack against a Web server (Section~\ref{sec:case1}).
We then provide a number of real-world settings in which \approach' prediction labels can be modified to achieve specific goals, for example predicting entire classes of attacks (Section~\ref{sec:case2}).

\subsection{Predicting events in a multi-step attack}
\label{sec:case1}

The first scenario where \approach' prediction capability can be leveraged is when facing multi-step, coordinated attacks, \ie a single attack involving multiple steps performed sequentially or in parallel by an attacker and resulting in multiple alerts being raised by the IPS. The difficulty of identifying such attacks originates from the fact that some of the intermediate steps of a multi-step attack can be considered benign when seen individually by an IPS engine. Moreover, most attacks observed in the wild are the result of automated scripts, which are essentially programmed to check for some precondition on the victim systems and subsequently trigger the adequate exploit(s). For instance, an attack might consists of the following steps: (i) run reconnaissance tasks if port \texttt{80/tcp} (HTTP) is open, (ii) trigger a list of exploits against the Web application framework, e.g., Apache and (iii) execute a list of exploits against other possible applications running on top of it. Therefore, we may not observe all steps of an attack on every victim system, depending on which branches of the attack scripts were executed. This variability of observed events across systems hinders the identification of the global multi-step attack.

To identify candidate multi-step attacks in our dataset of IPS events we used the following approach. For each event $e_i$ observed on any of the monitored machines we compute its frequency of occurrence across all machines. We then consider a candidate multi-step attack any sequence of events $e_i$ occurring at the same frequency with an error margin of $10$\% to capture the variability in attacks as explained above. We also set a \emph{support} threshold on the number of machines exhibiting that sequence so as to avoid a biased frequency obtained from too few samples. To uncover the case study presented here we empirically set this threshold to $1000$ machines. For network-sourced attack steps, we also extract the source IP address to determine the likelihood of the global event sequence to be generated by a single attacker.

We present the case of a multi-step attack captured by the IPS and which was successfully learned by the prediction tool. The attack consists of multiple attempts to exploit a Web server and the Web application running on top of it. First, the attacker checks the Web server software for several vulnerabilities. In this case, it quickly identifies the server as running Apache and then attempts to exploit several vulnerabilities, such as Struts-related vulnerabilities. It then switches to checking the presence of a Web application running on top and then fingerprinting it. In this step, the attacker triggers different exploits against known vulnerabilities in various Content Management Systems, such as Wordpress, Drupal, Joomla. Eventually, the attack appears to fail as the various steps are individually blocked by the IPS.

\begin{figure*}[t]
	\centering
    \begin{subfigure}[t]{0.67\textwidth}
        \raisebox{-\height}{\includegraphics[width=\linewidth]{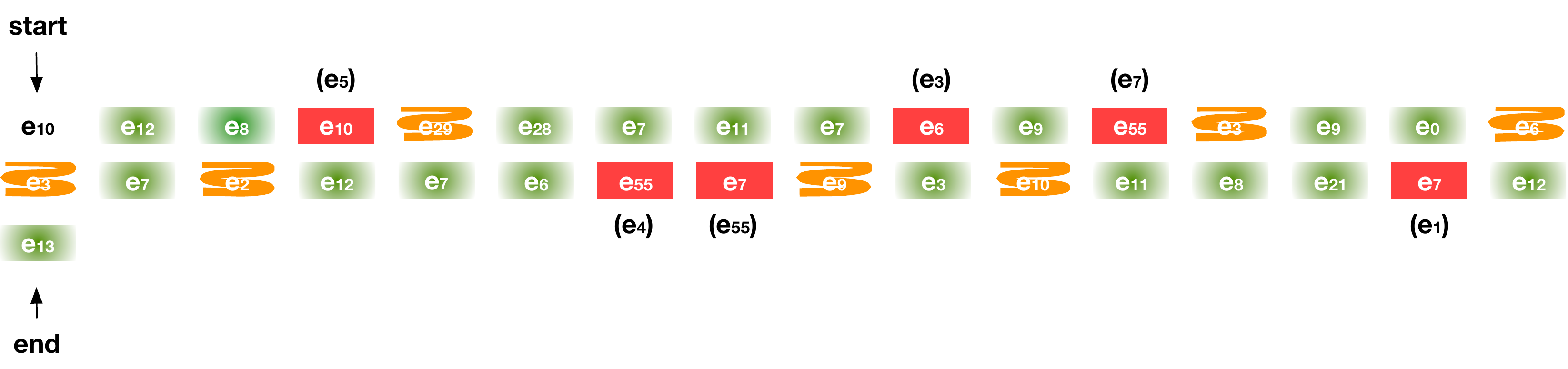}}
	 \caption{\approach prediction process in $machine_1$}
    \end{subfigure}
 
    \begin{subfigure}[t]{0.67\textwidth}
        \raisebox{-\height}{\includegraphics[width=\linewidth]{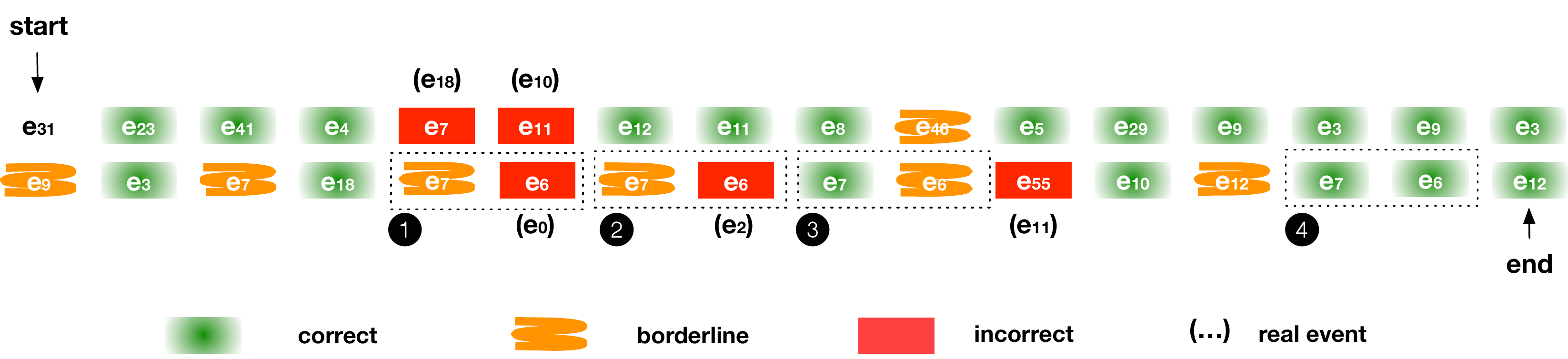}}
    \caption{\approach prediction process in $machine_2$}
    \end{subfigure}

	\caption{Step by step visualization of \approach prediction process in two real systems. \approach starts with event \textbf{$e_{10}$} and \textbf{$e_{31}$} respectively as the initial feed and predicts the upcoming security event step by step. The predictions are colored by their probabilistic scores, where green indicates \approach returns a correct prediction with probabilistic score greater than $0.5$, orange indicates \approach returns a correct event prediction with probabilistic score less than $0.5$ (but remains the largest probabilistic score), and red indicates a wrong prediction (the actual events are shown in parentheses in this case).}
	\label{fig:model_prediction_visualisation}
	\vspace{-0.3cm}
\end{figure*}

To be able to visualize the decision process and explain how \approach operates given the aforementioned multi-step attack, we feed \approach a list of security events from a machine that was under the coordinated attack. By putting \approach in this real-world environment, we are able to visualize how \approach predicts the upcoming events as illustrated in Figure~\ref{fig:model_prediction_visualisation}. Note that events \{$e_0$, ..., $e_{13}$\} belong to the coordinated attack.

The process operates as follows. Take $machine_1$ in Figure~\ref{fig:model_prediction_visualisation}a for example, \approach takes event $e_{10}$ as the initial feed and predicts the upcoming event $e_{12}$. It then verifies with the actual event to check if $e_{12}$ is the correct prediction. In our case, $e_{12}$ is the correct prediction with a confidence score higher than 0.5 (therefore $e_{12}$ in a green box in Figure~\ref{fig:model_prediction_visualisation}a). \approach automatically leverages both $e_{10}$ and $e_{12}$ as ``contextual information'' to enable its internal memory array cells to better predict the next event. The same step is repeated: $e_{8}$ is correctly predicted, \approach uses $e_{10}, e_{12}, ~\text{and}~ e_{8}$ to prepare its internal cells. In the case that \approach makes a wrong prediction, \eg it predicts $e_{10}$ instead of the actual observed event $e_5$ ($e_{10}$ is enclosed in a red box in Figure~\ref{fig:model_prediction_visualisation}a), \approach uses the actual event $e_5$ together with previous historical events. This enables \approach to stay on track with the observed events and predict events that are closely relating to those of the coordinated attack. This may lead \approach to incorrectly predict some random attacks the system experienced. For example, $e_{55}$ is a `PHP shell command execution' attack which was observed in the security event sequences, but not part of this particular coordinated attack. It is important to notice that \approach is able to correctly predict $e_{13}$ (an attack relating to CVE-2017-9805) that was not presented in the previous events, even thought its predecessors, such as $e_{12}$, $e_{3}$, $e_{10}$, appeared multiple times. We consider this a good example of \approach using long-term memory to carry out the correct prediction as detailed in Section~\ref{sec:seqlen}. It is also interesting to see how \approach adapts itself during its operation as shown in Figure~\ref{fig:model_prediction_visualisation}b (multiple $e_7$ and $e_6$ in dashed line rectangles). We can observe that \approach did not correctly predict $e_6$ twice when given $e_7$ (see \ding{182} and \ding{183}). Nevertheless, \approach is capable of leveraging the contextual information (\ie the actual observed events) to rectify its behavior. As we can observe in \ding{184}, \approach is able to make a borderline prediction and in \ding{185} \approach makes a confident and correct prediction of $e_6$.

To further exemplify \approach' sequential prediction capability in the above setup, we run it on 8 February (2018) test data using the model trained on 3 January (2018) and predict all upcoming events of 200 randomly selected  machines (this effectively generates 32,391 sequences due to the step-by-step setup) with a precision of 94\%, and against 8 December (2017) test data using a model trained on 4 November (2017) and obtain a precision of 80.89\%. These experimental results provide additional evidence of \approach' prediction capability in a real-world environment.

\subsection{Adjusting the prediction granularity}
\label{sec:case2}

The goal of \approach is primarily to accurately determine the next event that is going to occur on a given monitored system. In some cases multiple security events might share some common traits. For example, multiple IPS events can be used to describe different attacks against a particular software application, network protocol, etc. These shared traits can then be used to categorize such events. This categorization is specific to the security application that generated these events. Also, the categorization process undoubtedly results in a compressed and coarser-gained set of security events. In this section we describe several cases where such a categorization can be leveraged when the system fails to predict the exact security event but successfully predicts the exact traits, or categories of the attacks, such as the targeted network protocol and software application, or the attack type. To begin with, we extracted categories from the IPS signature labels and descriptions. These categories correspond to characteristics of attacks described by these signatures and are defined as follows. Whenever possible, we identify the \emph{verdict} of the signature, the \emph{severity} of the attack, the \emph{type of attack}, \eg remote command execution (RCE), SQL injection, etc, the \emph{targeted application}, if any, the \emph{targeted network protocol}, if any, and whether the attack exploits a particular \emph{CVE}. There is thus a one-to-many relationship between each signature and the categories it belongs to. We then uncover machines for which the categories of a mistakenly predicted event matches exactly the categories of the correct prediction. About $3.5$\% of failed prediction results exhibit this pattern.

For our first example we consider a machine that was targeted by the Shellshock exploits targeting the Unix shell BASH. Several vulnerabilities were uncovered in the context of these infamous attacks, namely CVE-2014-6271, CVE-2014-6277, CVE-2014-6278, CVE-2014-7169, CVE-2014-7186 and CVE-2014-7187. These six vulnerabilities translate into six IPS signatures. These signatures all belong to the same categories, which include (i) \texttt{block}, (ii) \texttt{high}, (iii) \texttt{RCE}, (iv) \texttt{bash} and (v) \texttt{CVE}. These categories mean that the exploit attempt is meant to be (i) \emph{blocked} because its potential security impact on the targeted machine is of (ii) \emph{high} severity. This verdict is explained by the fact that, if successful, the exploit would enable the attacker to perform a \emph{remote code execution (or RCE)} by exploiting a known \emph{vulnerability} (with an assigned CVE identifier) against the Unix shell \emph{BASH}. In this case study, a machine was targeted by several of the Shellshock exploits. After observing an attempt to exploit CVE-2014-6271, the system predicted another attempt to exploit the same vulnerability, instead of the correct prediction for CVE-2014-6278. While the event-level prediction result is wrong, the category-level prediction successfully identify an attempt to exploit a Shellshock vulnerability.

The second example is related to a machine that has apparently visited or have been triggered to visit a website distributing fake anti-virus software. Several IPS signatures have been defined to capture different aspects of these malicious websites, for instance, regular expressions matching specific HTML content, suspicious JavaScript code, etc. In this example, the system predicted that the machine would be redirected to a fake AV website containing a particular piece of malicious HTML code while in reality, the machine was redirected to a fake AV website containing a malicious piece of JavaScript code.

Additionally, we evaluate the performance of our security event forecast model in predicting if the upcoming event should be blocked or allowed, a relaxation as the aim is not to determine the exact event that will happen, but if it is a low-priority alarm (that is still allowed by the product we receive the data from) or if it is a high-priority one (that is blocked immediately). This is one of the essential tasks that a predictive intrusion prevention system needs to resolve. Our experiment shows that the proposed predictive model is able to achieve 88.9\% precision in predicting if the upcoming event should be blocked or allowed. This represents a 8\% precision increase comparing to the exact event prediction on the same day. Nevertheless, The added value of adjusting the prediction granularity obviously depends on the accuracy of the categorization and the expected level of granularity of the prediction. 

\section{Discussion} \label{sec:limitations}

\noindent \textbf{Limitations of \approach.} A recurrent neural network, broadly speaking, is a statistical model. The more the model ``sees'' (i.e., the more training data) the better the prediction performance is. For rare events, since the model does not have enough training samples, \approach may not correctly predict these rare intrusion attempts. Existing statistical and machine learning methods are yet to offer a satisfactory solution to this problem~\cite{weiss1998learning,wu2004learning, kwon2012unified}. It would also be interesting to understand whether the recent work by Kaiser \etal~\cite{kaiser2017learning} that makes deep models learn to remember rare words can be applied to predict rare intrusion attempts. 
DeepLog, a previously proposed system~\cite{du2017deeplog}, focused on anomaly detection in regulated environments, such as Hadoop and OpenStack, with limited variety of events. In such a specific log environment, DeepLog is able to use a small fraction of normal log entries to train and achieves good detection results. Nevertheless, DeepLog still requires a small fraction of normal log entries would generate enough representative events and patterns. Another limitation following rare events prediction is model retraining when new security events (\eg new signatures) are created. This retraining is inevitable because machine learning models can only recognize events they have been trained upon. Our experimental results (Section~\ref{sec:comparison}) show that \approach takes around 10 hours to retrain and can be redeployed in a timely fashion in a real-world scenario.
As mentioned in Section \ref{sec:seqlen} the nodes that are activated in an LSTM are not easy to examine. For this reason we cannot guarantee that the system does not take into consideration spurious correlations. At the same time we tried to limit this issue by extensively evaluate \approach over a large amount of data and in different settings.

\noindent \textbf{Data limitations.} For its operation, \approach relies on a dataset of pre-labeled security events. An inherent limitation of this type of data is that an event can be labeled only if it  belongs to a known attack class. If, for example, a new zero-day vulnerability started being exploited in the wild, this would not be reflected in the data until a signature is created for it. To reduce the window between when an attack is being run and when it starts being detected by an intrusion prevention system security companies typically use threat intelligence systems and employ human specialists to analyze unlabeled data looking for new attack trends.

\noindent \textbf{\approach performance.} Sections~\ref{sec:evaluation} and~\ref{sec:case_studies} show the effectiveness of the system. The prediction of a security event in such a complicated environment is an important challenge. \approach shows the ability of effectively tackling this challenge, showing stability, even when the training set is months older than the test set, and robustness to noise while detecting multistage attacks. We evaluated \approach over different time periods to thoroughly prove its qualities; as we discussed, the system may need retraining only in case the data presents radical changes, while its precision does not decrease quickly if the training set is older than the test set. The system can support different dimensions of the training set as it has been tested using one day or one week of data. The differences are minimal: performance is extremely similar, but weekly training seems slightly more robust to anomalies on a specific day of data. However, weekly training sets require more time to build the model. Sections~\ref{sec:seqlen} and~\ref{sec:case_studies} show how long-term memory and noise filtering are both important factors in the precision of the neural networks, explaining why the baseline methods used in Section~\ref{sec:comparison} are less precise than \approach.

\noindent \textbf{Deployment.} The architecture of \approach enables it to be reasonably flexible in terms of real-world deployment. \approach can be deployed for each endpoint to proactively defend against attacks as we can see in Section~\ref{sec:case_studies}. At the same time, \approach can be tailored to protect an enterprise by training with the data coming from that enterprise only and thus better deal with the attacks targeting that enterprise. Note that \approach' predictive model trained using one day of data is about 31MB. It can be easily pushed to the endpoints with limited network footprint. Together with the mobile TensorFlow library, it is practically feasible for \approach to protect mobile/embedded devices by training with security event data coming from those devices only. For example, \approach can be trained using the data collected by smart routers with an IPS installed and deployed in these routers to protect smart home environments.

\noindent \textbf{Evasion.} \approach may be subject to evasion techniques from malicious agents. A vulnerability of deep learning systems is that while the system is classifying samples, it adapts its rules. Therefore, it may be subject to poisoning attacks from a criminal who influences the decision rules using fake actions before attacking the victim. However, to achieve such evasion, the attackers must apply such fake actions at a massive scale and target thousands of machines.
A technique that could be used by adversaries is mimicry attacks, \ie evading security systems by injecting many irrelevant events to cover the alerts generated by a real attack. We argue that \approach has the potential to be resistant to these attacks. Indeed, we have seen in the case studies that \approach is able to filter out the noise from the sequences of events observed on the machines, and detect the important events correctly. An interesting future work would be to be able to quantify the amount of events necessary to evade systems like \approach.
Zero day attacks may be difficult to detect: a zero day attack may replicate known sequences of actions to exploit new vulnerabilities, but that would still be detected; when the zero day is applying a new kind of multi-step attack that has a different sequence of events, it may not be detected.

\section{Related Work} \label{sec:relatedwork}

In this section, we broadly reviewed previous literature in both forecasting security events and deep learning, especially recurrent neural networks (RNNs), applications in security analytics.  

\subsection{Security Event Forecast}
\noindent \textbf{System-level security event forecast.} Soska \etal~\cite{soska2014automatically} proposed a general approach to predict with high certainty if a given website will become malicious in the future. The core idea of this study is building a list of features that best characterize a website, including traffic statistics, filesystem structure, webpage structure \& contents and statistics heuristic of dynamic features (\eg contents). These features are later used to train an ensemble of C4.5 classifiers. This model is able to achieve  operate with 66\% true positives and 17\% false positives given one-year data. Bilge \etal~\cite{leyla2017riskteller} proposed a system that analyzes binary appearance logs of machines to predict which machines are at risk of infection. The study extracts 89 features from individual machine's file appearance logs to produce machine profile, then leverages both supervised and semi-supervised methods to predict which machines are at risk. In terms of machine wise infection prediction, RiskTeller can predict 95\% of the to-be-infected machines with only 5\% false positives; regarding enterprise-wise infection prediction, Riskteller can, on average, achieve 61\% TPR with 5\% FPR.

\noindent \textbf{Organization-level security event forecast.} Liu \etal~\cite{liu2015cloudy} explored the effectiveness of forecasting security incidents. This study collected 258 externally measurable features about an organization's network covering two main categories: mismanagement symptoms (e.g., misconfigured DNS) and malicious activities (e.g., spam, scanning activities originated from this organization's network). Based on the data, the study trained and tested a Random Forest classifier on these features, and are able to achieve with 90\% True Positive (TP) rate, 10\% False Positive (FP) rate and an overall accuracy of 90\% in forecasting security incidents. Liu \etal~\cite{liu2015predicting} carried out a follow-up study on externally observed malicious activities associated with network entities (e.g., spam, phishing, malicious attacks). It further proved that when viewed collectively, these malicious activities are stable indicators of the general cleanness of a network and can be leveraged to build predictive models (e.g., using SVM). The study extracts three features: intensity, duration, and frequency, from this activity data. It later trained a SVM model using these features and achieved reasonably good prediction performance over a forecasting window of a few months achieving 62\% true positive rate with 20\% false positive rate.

\noindent \textbf{Cyber-level security event forecast.} Sabottke \etal~\cite{sabottke2015vulnerability} conducted a quantitative and qualitative exploration of the vulnerability-related information disseminated on Twitter. Based on the analytical results, the study designed a Twitter-based exploit detector, building on top 4 categories of features (Twitter Text, Twitter Statistics, CVSS Information and Database Information), for early detection of real-world exploits. This classifier achieves precision and recall higher than 80\% for predicting the existence of private proof-of-concept exploits when only the vulnerabilities disclosed in Microsoft products and by using Microsoft's Exploitability Index are considered.

\subsection{Recurrent Neural Network Applications in Security Research}
\noindent \textbf{Binary Analysis.} Shin \etal~\cite{shin2015recognizing} leveraged recurrent neural networks (RNN) to identify functions (\eg function boundaries, and general function identification) in binaries. For each training epoch, the RNN model is trained on $N$ examples (an example refers to a fixed-length sequence of bytes). The authors used one-hot encoding to convert each byte in a given example into a 256-vector, and associated a function start/end indicator with each byte (i.e., a 256-vector). Once the model is trained, it effectively serves as a binary classifier and outputs a decision for that byte as to whether it begins a function or not. The authors consequently combine the predictions from each model using simple heuristic rules to achieve aforementioned function identification tasks. It is claimed that this system is capable of halving the error rate on six out of eight benchmarks, and performs comparably on the remaining two. Chua \etal~\cite{chua2017neural} presents {\sc eklavya}, a RNN-based engine to recover function types (\eg identifying the number and primitive types of the arguments of a function) from x86/x64 machine code of a given function without prior knowledge of the compiler or the instruction set. On the condition that the boundaries of given functions are known, {\sc eklavya} developed two primary modules - instruction embedding module and argument recovery module - to recover argument counts and types from binaries. The instruction embedding module takes a stream of functions as input and outputs a 256-vector representation of each instructions. After the instructions are represented as vectors, argument recovery module uses these sequences of vectors as training data and trains four RNNs for four tasks relating to function types recovery. The authors reported accuracy of around 84\% and 81\% for function argument count and type recovery tasks respectively.

\noindent \textbf{Anomaly Detection.} Du \etal~\cite{du2017deeplog} proposed DeepLog, a deep neural network model utilizing Long Short-Term Memory (LSTM), to learn a system's log patterns (e.g., log key patterns and corresponding parameter value patterns) from normal execution. At its detection stage, DeepLog uses both the log key and parameter value anomaly detection models to identify abnormal log entries. Its workflow model provides semantic information for users to diagnose a detect anomaly. The author reported that DeepLog outperformed other existing log-based anomaly detection methods achieving a $F$-measure of 96\% in HDFS data and a $F$-measure of 98\% in OpenStack data.

\noindent \textbf{Password Attack.} Melicher \etal~\cite{melicher2016fast} used artificial neural networks to model text passwords' resistance to guessing attacks and explore how different architectures and training methods impact neural networks' guessing effectiveness. The authors demonstrated that neural networks guessed 70\% of 4class8 (2,997 passwords that must contain all four character classes and be at least eight characters) passwords by $10^{15}$ guesses, while the next best performing guessing method (Markov models) guesses 57\%.

\noindent \textbf{Malware Classification.} Pascanu \etal~\cite{pascanu2015malware} model malware API calls as a sequence and use a recurrent model trained to predict next API call, and use the hidden state of the model (that encodes the history of past events) as the fixed-length feature vector that is given to a separate classifier (logistic regression or MLP) to classify malware.

The closest work to this paper is DeepLog~\cite{du2017deeplog}. However, DeepLog focused on anomaly detection in regulated environment such as Hadoop and OpenStack with limited variety of events (e.g., 29 events in Hadoop environment and 40 events in OpenStack). In such a very specific log environment, DeepLog was able to use a small fraction of normal log entries to train and achieve good detection results. Our work aims at understanding multi-steps coordinated attacks in a noisy environment with a wide variety of events (\ie 4,495 unique events in our dataset) and prediction in this setup is a far harder problem comparing to DeepLog. Additionally DeepLog considered an event \emph{abnormal} if such event is \emph{not} with top-$g$ probabilities to appear next. Our work does not employ this relaxed prediction criteria and focuses on the accurate prediction of the upcoming security event (out of 4,495 possible events) for a given system. 

\section{Conclusions} \label{sec:conclusion}
This paper presented \approach, a system for the prediction of security events. We evaluated it using an extensive dataset of alarms from an intrusion prevention system from a major security firm product. \approach reaches a high precision for such a complex problem, showing stable results even when the model is trained months before the application to a test set.

\begin{acks}
We wish to thank the anonymous reviewers for their feedback and our shepherd Zhou Li for his help in improving our paper.
Enrico Mariconti was supported by the EPSRC under grant 1490017.
\end{acks}

\bibliographystyle{ACM-Reference-Format}
\bibliography{bibliography}


\balance

\end{document}